\documentclass{aa}
\usepackage[varg]{txfonts}
\usepackage{amsmath,amstext}

\usepackage{graphicx}
\usepackage{epstopdf}
\usepackage{float}
\usepackage{array}
\usepackage{mathtools}
\usepackage{booktabs}
\usepackage{subfigure}
\usepackage{url}
\usepackage{helvet}
\usepackage{tabularx}
\usepackage{multirow}
\usepackage{natbib}
\usepackage[flushleft]{threeparttable}

\usepackage{lscape}
\usepackage{pdflscape}
\usepackage{longtable}
\usepackage{wasysym}
\usepackage{relsize}
\usepackage{color}
\usepackage{breqn}
\usepackage[breaklinks=true]{hyperref} 
\definecolor{darkblue}{rgb}{0.0,0.0,0.4}
\definecolor{darkgreen}{rgb}{0.0,0.4,0.0}
\definecolor{violet}{rgb}{0.5,0.,0.5}
\hypersetup{colorlinks,breaklinks, linkcolor=darkblue,urlcolor=darkblue, anchorcolor=darkblue,citecolor=darkblue}
\def\kms{km s$^{-1}$}         
\def\ms{\hbox{m s$^{-1}$}}         
\def\gcm3{\hbox{g cm$^{-3}$}}       
\def\Msun{\hbox{$\mathrm{M}_{\astrosun}$}}             
\def\Rsun{\hbox{$\mathrm{R}_{\astrosun}$}}

\def\Mearth{\hbox{$\mathrm{M}_{\oplus}$}}
\def\Rearth{\hbox{$\mathrm{R}_{\oplus}$}}
\def\degr{\hbox{$^\circ$}}
\def\teff{T$_{\rm eff}$}
\def\logg{log~{\it g}}
\def\met{[Fe/H]}

\def\figw{\columnwidth}

\bibpunct{(}{)}{;}{a}{}{,} 
\begin{document}

\title{K2-265 b: A Transiting Rocky Super-Earth}
\author{K.~W.~F.~ Lam\inst{\ref{warwick}\thanks{Now at Zentrum f\"ur Astronomie und Astrophysik, Technische Universit\"at Berlin, Hardenbergstr. 36, 10623 Berlin, Germany; email: k.lam@tu-berlin.de}}    
       \and    A.~Santerne\inst{\ref{LAM}}  
       \and    S.~G.~Sousa\inst{\ref{IA}}   
       \and    A.~Vigan\inst{\ref{LAM}}
       \and    D.~J.~Armstrong\inst{\ref{warwick}\fnmsep\ref{CEH}}
       \and    S.~C.~C.~Barros\inst{\ref{IA}}
       \and    B.~Brugger\inst{\ref{LAM}}
       \and    \\V.~Adibekyan\inst{\ref{IA}}
       \and    J.-M.~Almenara\inst{\ref{geneva}}
       \and    E.~Delgado~Mena\inst{\ref{IA}}
       \and    X.~Dumusque\inst{\ref{geneva}}
       \and    D.~Barrado\inst{\ref{CAB}}
       \and    D.~Bayliss\inst{\ref{warwick}}
       \and    A.~S.~Bonomo\inst{\ref{Torino}}
       \and    F.~Bouchy\inst{\ref{geneva}}
       \and    D.~J.~A.~Brown\inst{\ref{warwick}\fnmsep\ref{CEH}}
       \and    D.~Ciardi\inst{\ref{caltech}} 
       \and    M.~Deleuil\inst{\ref{LAM}}       
       \and    O.~Demangeon\inst{\ref{IA}}
       \and    F.~Faedi\inst{\ref{catania}\fnmsep\ref{warwick}}        
       \and    E.~Foxell\inst{\ref{warwick}}    
       \and    J.~A.~G.~Jackman\inst{\ref{warwick}\fnmsep\ref{CEH}} 
       \and   G.~W.~King\inst{\ref{warwick}}    
       \and   J.~Kirk\inst{\ref{warwick}} 
       \and    R.~Ligi\inst{\ref{LAM}} 
       \and    J.~Lillo-Box\inst{\ref{ESO}}
       \and    T.~Lopez\inst{\ref{LAM}}
       \and    C.~Lovis\inst{\ref{geneva}}
       \and    T.~Louden\inst{\ref{warwick}\fnmsep\ref{CEH}}
       \and    L.~D.~Nielsen\inst{\ref{geneva}}
       \and    J.~McCormac\inst{\ref{warwick}\fnmsep\ref{CEH}}
       \and    O.~Mousis\inst{\ref{LAM}}      
       \and    H.~P.~Osborn\inst{\ref{LAM}}
       \and    D.~Pollacco\inst{\ref{warwick}\fnmsep\ref{CEH}}   
       \and    N.~C.~Santos\inst{\ref{IA}\fnmsep\ref{UPorto}}    
       \and    S.~Udry\inst{\ref{geneva}}
       \and    P.~J.~Wheatley\inst{\ref{warwick}\fnmsep\ref{CEH}}
       }      

   \institute{
  Department of Physics, University of Warwick, Gibbet Hill Road, Coventry, CV4 7AL, UK\label{warwick}
 \and Aix Marseille Univ, CNRS, CNES, LAM, Marseille, France\label{LAM}
  \and Instituto de Astrof\'isica e Ci\^{e}ncias do Espa\c co, Universidade do Porto, CAUP, Rua das Estrelas, 4150-762 Porto, Portugal\label{IA}
  \and Centre for Exoplanets and Habitability, University of Warwick, Gibbet Hill Road, Coventry CV4 7AL, UK\label{CEH}
 \and Observatoire Astronomique de l'Universit\'e de Gen\`eve, 51 Chemin des Maillettes, 1290 Versoix, Switzerland\label{geneva}
 \and Depto. de Astrof\'isica, Centro de Astrobiolog\'ia (CSIC-INTA), ESAC campus 28692 Villanueva de la Ca\~nada (Madrid), Spain\label{CAB}
 \and INAF -- Osservatorio Astrofisico di Torino, Strada Osservatorio 20, I-10025, Pino Torinese (TO), Italy\label{Torino}
  \and Caltech/IPAC-NASA Exoplanet Science Institute, 770 S. Wilson Ave, Pasadena, CA 91106, USA\label{caltech}
 \and  INAF - Osservatorio Astrofisica di Catania, via S. Sofia 78, 95123, Catania Italy \label{catania}
 \and European Southern Observatory (ESO), Alonso de Cordova 3107, Vitacura, Casilla 19001, Santiago de Chile, Chile\label{ESO}
 \and Departamento de F\'isica e Astronomia, Faculdade de Ciencias, Universidade do Porto, Rua Campo Alegre, 4169-007 Porto, Portugal\label{UPorto}
           }

\date{Received date / Accepted date }

\abstract {We report the discovery of the super-Earth K2-265 b detected with K2 photometry. The planet orbits a bright ($\rm V_{mag} = 11.1$) star of spectral type G8V with a period of $2.37$ days. We obtained high-precision follow-up radial velocity measurements from HARPS, and the joint Bayesian analysis showed that K2-265 b has a radius of $\rm 1.71 \pm 0.11 ~R_{\oplus}$ and a mass of $\rm 6.54 \pm 0.84 ~M_{\oplus}$, corresponding to a bulk density of $7.1 \pm 1.8\ \rm g\ cm^{-3}$. Composition analysis of the planet reveals an Earth-like, rocky interior, with a rock mass fraction of $\sim80\%$. The short orbital period and small radius of the planet puts it below the lower limit of the photoevaporation gap, where the envelope of the planet could have eroded due to strong stellar irradiation, leaving behind an exposed core. Knowledge of the planet core composition allows us to infer the possible formation and evolution mechanism responsible for its current physical parameters.}


\titlerunning{K2-265 b}
\authorrunning{Lam et al.}

\maketitle

\section{Introduction}
\hspace{0.5cm}Exoplanetary discovery has widened our perspective and knowledge of planetary science in the past two decades. The space-based mission \textit{Kepler} used transit photometry to detect and characterise exoplanets (\citealt{2010Sci...327..977B,2011ApJ...736...19B,2010ApJ...713L..79K}), with one of its key objectives being the determination of the frequency of terrestrial planets in the habitable zones of stars. From their sample of over 4000 transiting planet candidates, it was revealed that small planets ($R_P < 4.0 R_{\oplus}$) are by far the most common in our Galaxy (\citealt{2012ApJS..201...15H,2013ApJS..204...24B,2013ApJ...767...95D,2013PNAS..11019273P}), a result that is also supported by radial-velocity surveys (e.g. \citealt{2013A&A...549A.109B,2011arXiv1109.2497M}). While the \textit{Kepler} sample provided an insight into the planet occurrence rate (e.g. \citealt{2014PNAS..11112647B}), only a few dozen host stars were bright enough for follow-up characterisation. With the loss of two reaction wheels on the \textit{Kepler} spacecraft, the \textit{K2} mission was adopted to extend the transiting exoplanet discoveries \citep{2014PASP..126..398H}. \textit{K2} has observed nineteen fields so far, and supplied precise photometry of approximately $20,000$ bright stars per campaign. This has yielded hundreds of transiting planet candidates (e.g. \citealt{2016ApJS..222...14V,2016A&A...594A.100B,2016MNRAS.461.3399P}), over 300 of which have been statistically validated (e.g. \citealt{2015ApJ...809...25M,2015MNRAS.454.4267B,2016ApJS..226....7C}). 

  Super-Earths are absent in our own Solar system. Therefore, they are of particular interest in the study of planet formation and evolution. To probe the formation histories of these small planets, it is necessary to derive the planetary masses and radii with precision better than a few percent in order to differentiate their internal compositions in the context of planet evolution models (e.g. \citealt{2013PASP..125..227Z,2017ApJ...850...93B}). Recent theories have proposed a distinct transition in the composition of small exoplanets (\citealt{2014ApJ...783L...6W,2015ApJ...801...41R}). Planets with $\rm R_P \lesssim 1.6 ~R_{\oplus}$ typically have high densities and are predominantly rocky. On the other hand, planets with larger radii typically have lower densities and possess extended H/He envelopes. In fact, planets such as Kepler-10~b ($\rm R_P = 1.42 \pm 0.03 ~R_{\oplus}$, $ \rho_{P} = 8.8 \pm 2.5\ \rm g\ cm^{-3}$; \citealt{2011ApJ...729...27B}), LHS1140~b ($\rm R_P = 1.43 \pm 0.10 ~R_{\oplus}$, $\rho_P = 12.5 \pm 3.4\ \rm g\ cm^{-3}$; \citealt{2017Natur.544..333D}), Kepler-20~b ($\rm R_P = 1.87 \pm 0.05 ~R_{\oplus}$, $ \rho_P = 8.2 \pm 1.4\ \rm g\ cm^{-3}$; \citealt{2016AJ....152..160B}), and K2-38~b ($\rm R_P = 1.55 \pm 0.02 ~R_{\oplus}$, $ \rho_P = 17.5  \pm 7.35\ \rm g\ cm^{-3}$; \citealt{2016ApJ...827...78S}) all have densities higher than that of the Earth ($\rho_{\oplus} = 5.5 \rm g\ cm^{-3}$) and compositions consistent with a rocky world, whereas low density planets such as GJ~1214~b ($\rm R_P = 2.68 \pm 0.13 ~R_{\oplus}$, $\rho_P = 1.87  \pm 0.40\ \rm g\ cm^{-3}$; \citealt{2009Natur.462..891C}), the Kepler-11 system ($\rm R_P = 1.97$--$4.52 ~R_{\oplus}$, $\rho_P = 0.5$--$3.1\ \rm g\ cm^{-3}$; \citealt{2011Natur.470...53L}), and HIP~116454~b ($\rm R_P = 2.53 \pm 0.18 ~R_{\oplus}$, $ \rho_P = 4.17  \pm 1.08\ \rm g\ cm^{-3}$; \citealt{2015ApJ...800...59V}) have solid cores, and substantial gaseous envelopes.

Recent efforts by the California-\textit{Kepler} Survey (CKS) (\citealt{2017AJ....154..108J,2017AJ....154..109F}) have refined the physical characteristics of \textit{Kepler} short-period planets ($\rm P < 100 ~days$) and their host stars for an in-depth study of the planet size distribution. Their results show a significant lack of planets with sizes between $1.5\ \rm R_{\oplus}$ and $2.0\ \rm R_{\oplus}$. The gap in the radius distribution can be explained by the `photoevaporation' model (\citealt{2013ApJ...775..105O,2013ApJ...776....2L}), where the gaseous envelopes of planets are stripped away as a result of exposure to high incident flux from their host stars. The CKS also highlighted the importance of obtaining precise measurements of planet masses and radii in order to perform statistically significant studies of the radius distribution.

In this paper, we report the detection of a $2.37$-day transiting super-Earth, K2-265 b. \textit{K2} photometry and HARPS radial velocity measurements were used to constrain the radius and mass measurements of this planet with a precision of 6\% and 13\%, respectively. In Section \ref{observations}, we describe the observations made from \textit{K2}, data reduction, and spectroscopic follow up. Our analyses and results are presented in Section \ref{results}, and we conclude the paper with a summary and discussion in Section \ref{conclusion}.

\section{Observations\label{observations}}

	\subsection{K2 Photometry\label{photometry}}
	\hspace{0.5cm}
    
    K2-265 was observed during \textit{K2} Campaign 3 in long cadence mode. The photometry was obtained between November 2014 and January 2015. The target was independently flagged as a candidate from two transit searches; the first made use of the POLAR pipeline \citep{2016A&A...594A.100B}, and the second used the methods described in \cite{2015A&A...579A..19A} and \cite{2015A&A...582A..33A}, where human input was involved to identify high priority candidates. K2-265 was also independently identified as a planet-hosting candidate by other search algorithms \citep{2016ApJS..222...14V, 2016ApJS..226....7C,2018AJ....155..136M}.
    
\begin{figure*}[htbp!]
	\centering
	    \includegraphics[width=\textwidth]{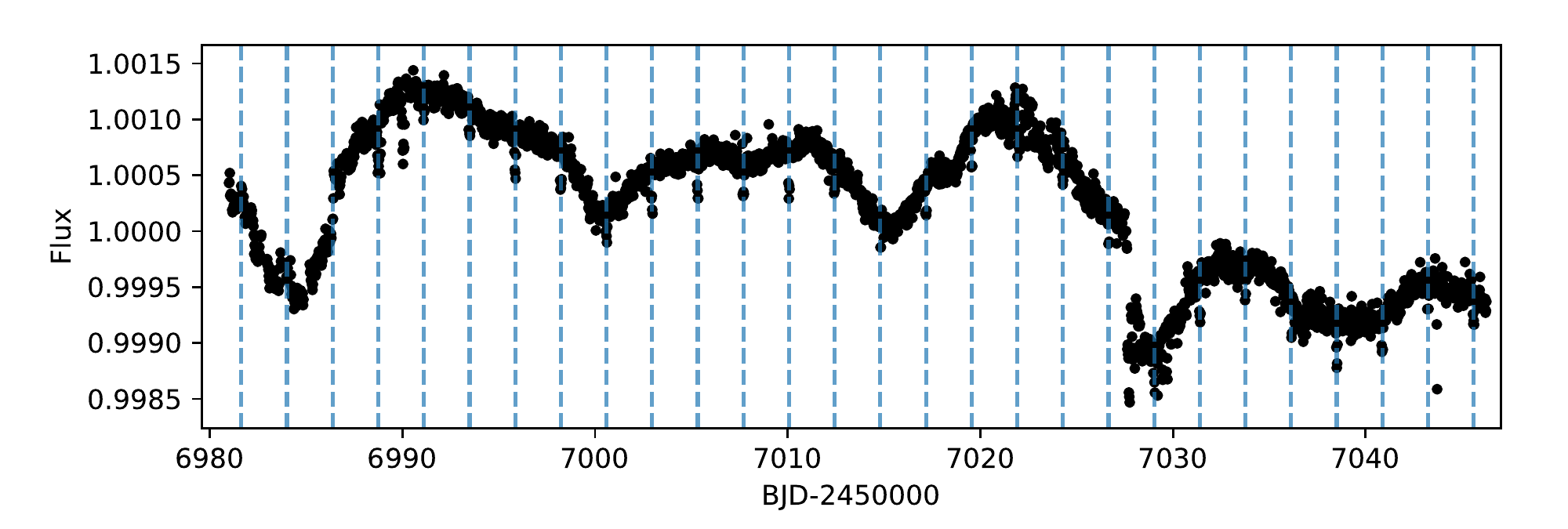}
	    \caption{\label{k2lc}The detrended \textit{K2} transit light curve of K2-265 with positions of transits marked with blue dashed lines. }
	\end{figure*}

    The \textit{K2} lightcurve generated from the POLAR pipeline \citep{2016A&A...594A.100B} has less white noise than that of \cite{2015A&A...579A..19A,2015A&A...582A..33A}, hence the former was used in the planetary system analysis. The POLAR pipeline is summarised as follows: 
The \textit{K2} pixel data was downloaded from the Mikulski Archive for Space Telescopes (MAST)\footnote{\url{http://archive.stsci.edu/kepler/data_search/search.php}}. The photometric data was extracted using the adapted CoRoT imagette pipeline \cite{2014A&A...569A..74B} which uses an optimal aperture for the photometric extraction. In this case, the optimal aperture was found to be close to circular and comprised of 44 pixels. The Modified Moment Method developed by \citet{1989AJ.....97.1227S} was used to determine the centroid positions for systematic corrections. Flux and position variations of the star on the CCD can lead to systematics in the data. These were corrected following the self-flat-fielding method similar to \citet{2014PASP..126..948V}. Figure \ref{k2lc} shows the final extracted lightcurve, and Table \ref{PhotometricProperties} gives the photometric properties of K2-265.
    
	\begin{figure}[htbp!]
	\centering
	    \includegraphics[width=0.5\textwidth]{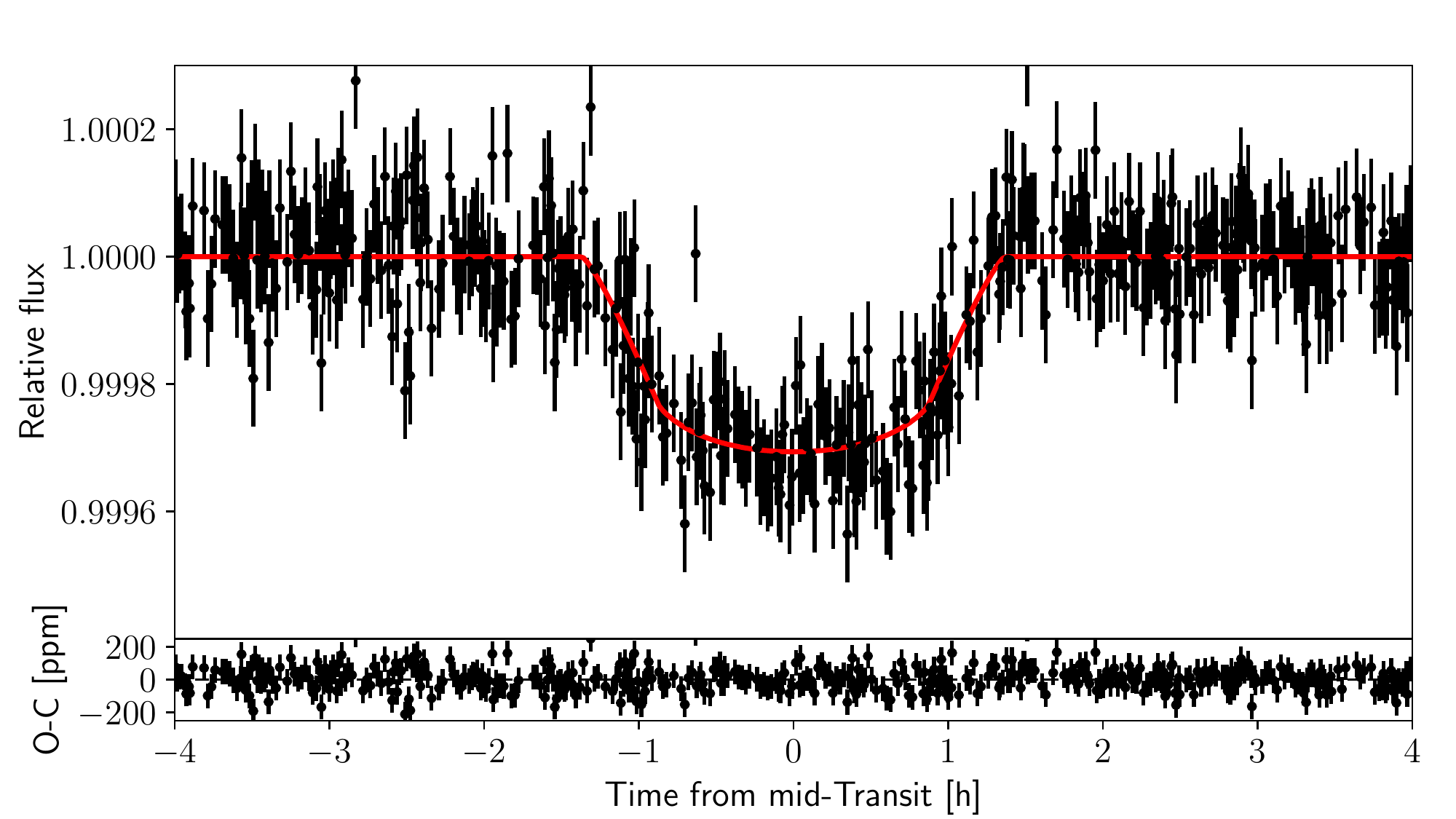}
	    \caption{\label{k2foldedlc}The phase-folded \textit{K2} lightcurve of K2-265 is shown in the top panel with the ephemeris from our analysis. The best-fit transit model is plotted in a red solid line, and the residuals of the fit are plotted in the bottom panel.} 
	\end{figure}

	\begin{table}
  \begin{threeparttable}
    \caption{\label{PhotometricProperties}Properties of K2-265. K2-265 has a nearby bound companion (see text for detailed description), hence values presented in this table are for the blended photometry. The photometric magnitudes listed were used in deriving the SED as described in Section \ref{MCMC}.}
     \begin{tabular}{lcc}
        \toprule
        Parameter & Value and uncertainty & Source\\
        \midrule
		K2 Campaign & 3 & a\\
		EPIC & 206011496 & a\\
		2MASS ID & 2MASS J22480755$-$1429407  & b\\
		RA(J2000) & 22:48:07.56 & c\\
		Dec(J2000) & $-$14:29:40.84  & c\\ 
		$\mu_{RA}$ (mas/yr) & $30.20 \pm 0.09$  & c\\
		$\mu_{DEC}$ (mas/yr) & $-23.34 \pm 0.06$ & c\\
		Parallax (mas) & $7.18 \pm 0.05$ & c\\
		\hline
		\multicolumn{3}{l}{\textit{Photometric magnitudes}}	\\
		Kp & $10.92$ & a\\ 
		Gaia G & $10.928$ & c\\ 
		Johnson B & $11.845\pm0.029$ & d\\
		Johnson V & $11.102\pm0.037$ & d\\
		Sloan g$^{\prime}$ & $11.419\pm0.042$ & d\\
		Sloan r$^{\prime}$ & $10.879\pm0.047$ & d\\
		Sloan i$^{\prime}$ & $10.689\pm0.084$ & d\\
		2-MASS J & $9.726\pm0.026$ & b\\
		2-MASS H & $9.312\pm0.022$ & b\\
		2-MASS Ks & $9.259\pm0.027$ & b\\
		WISE W1 & $9.178\pm0.022$ & e\\
		WISE W2 & $9.213\pm0.020$ & e\\
		WISE W3 & $9.162\pm0.040$ & e\\
        \bottomrule
     \end{tabular}
    \begin{tablenotes}
      \small
      \item a. EXOFOP-K2: \url{https://exofop.ipac.caltech.edu/k2/}
      \item b. The Two Micron All Sky Survey (2MASS)
      \item c. \textit{Gaia} DR2
      \item d. The AAVSO Photometric All-Sky Survey (APASS)
      \item e. AllWISE
    \end{tablenotes}
  \end{threeparttable}
\end{table}

		\subsection{Spectroscopic Follow Up\label{RV}}
\begin{figure*}[htbp!]
	\centering
	    \includegraphics[width=\textwidth]{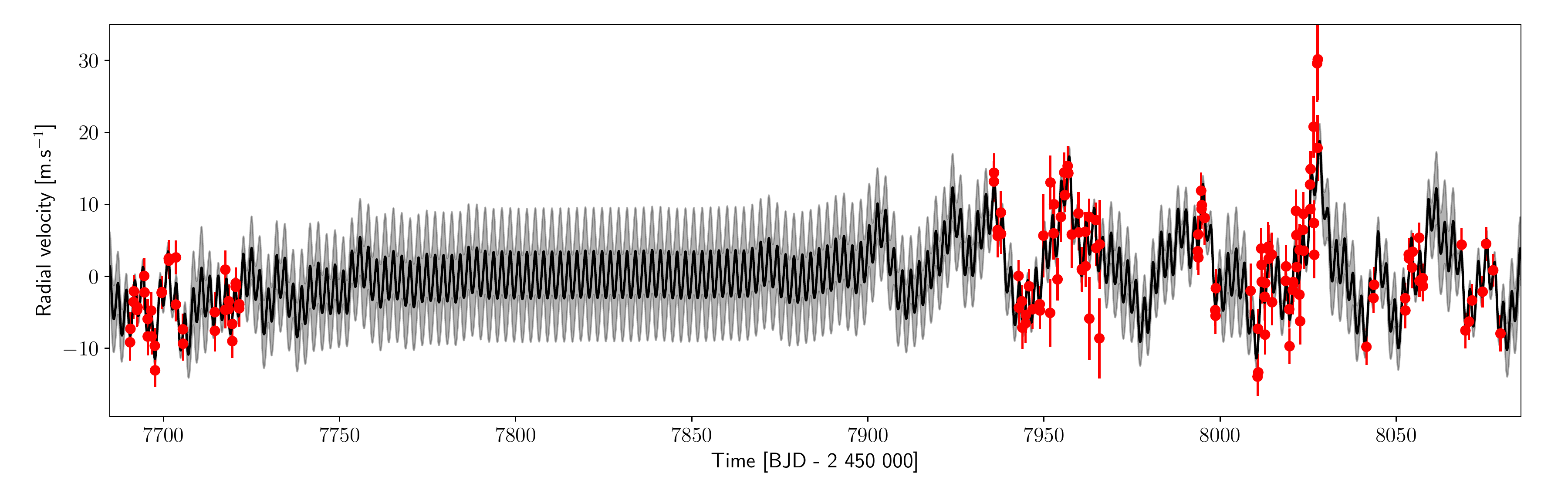}
	    \caption{\label{RVtimeseries}Time-series HARPS radial velocity measurements (red circles) of K2-265. The best-fit Keplerian orbit of K2-265 b is plotted in black. The stellar activity is fitted with a Gaussian process. The grey region show the 1-$\sigma$ confidence interval of the Gaussian process.}
	\end{figure*}

\begin{figure}[htbp!]
	    \includegraphics[width=0.5\textwidth]{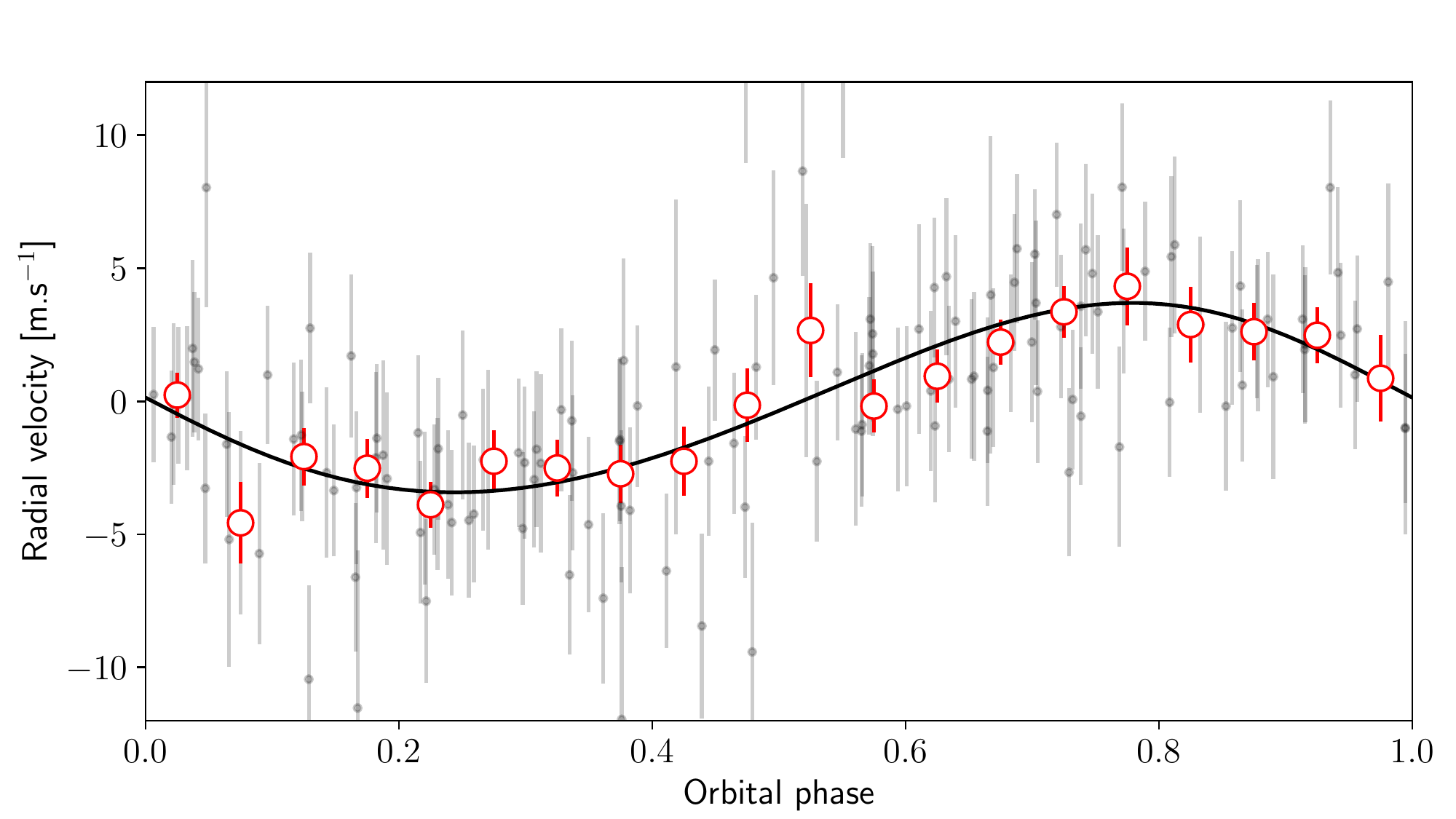}
	    \caption{\label{RVfold} \textit{Upper panel}: Phase-folded HARPS radial velocity measurements (black circles) of K2-265 as a function of the orbital phase. The black solid line is the best-fit RV curve. The binned RV measurements are denoted as red open circles.}
	\end{figure}

	\hspace{0.5cm}We obtained radial velocity (RV) measurements of K2-265 with the HARPS spectrograph ($\rm R\sim110,000$), mounted on the $3.6$ m Telescope at ESO La Silla Observatory \citep{2003Msngr.114...20M}. A total of 153 observations were made between 2016 October 29 and 2017 November 22 as part of the ESO-K2 large programme\footnote{Based on observations made with ESO Telescopes at the La Silla Paranal Observatory under programme ID 198.C-0169.}. An exposure time of 1800\,s was used for each observation, giving a signal-to-noise ratio of about 50 per pixel at 5500\,\AA. The data were reduced using the HARPS pipeline \citep{1996A&AS..119..373B}. RV measurements were computed with the weighted cross-correlation function (CCF) method using a G2V template (\citealt{1996A&AS..119..373B,2002Msngr.110....9P}), and the uncertainties in the RVs were estimated as described in \citet{2001A&A...374..733B}. The line bisector (BIS), and the full width half maximum (FWHM) were measured using the methods of \citet{2011A&A...528A...4B} and \citet{2015MNRAS.451.2337S}. 
Ten observations that were obtained when the target was close to a bright Moon exhibit a significant anomaly in their FWHM, up to 500 \ms. We removed these data completely from the analyses described in the later sections. The remaining 143 RV measurements and their associated uncertainties are reported in Table.~\ref{RVtable}. The time-series RVs and the phase-folded RVs of K2-265 are shown in Figures \ref{RVtimeseries} and \ref{RVfold} respectively. Following the calibrations of \cite{1984ApJ...287..769N}, we derived the activity index of $\log R'_{HK} = -4.90 \pm 0.12$. The activity index is used in Sect.~\ref{K2rotation} to derive the stellar rotational period.
	
	\subsection{Direct imaging observations\label{astrometry}}
	Shallow imaging observations were obtained with the NIRC2 instrument at Keck on 2015-08-04 in the narrow-band $Br_{\gamma}$ filter at 2.169~$\mu$m (programme N151N2, PI: Ciardi). Several images were acquired with a dithering pattern on-sky and they were simply realigned and median-combined. In the combined image, a candidate companion was clearly detected at close separation from the star. Figure \ref{KECKAO} shows the K-bank Keck AO image of K2-265 and the near-by companion, where the contrast of the objects is measured to be $\Delta \rm mag = 8.12$ in the K-band. The relative astrometry of the candidate was estimated using a simple Gaussian fitting on both the star and the candidate. The error on the measurement is conservatively estimated to $\sim$0.5~pixel, i.e. $\sim$5~mas. The relative Keck astrometry was derived following methods described in \cite{Vigan2016}, and the following parameters were obtained:  $\Delta\alpha = -910 \pm 5$ mas, $\Delta\delta = -363 \pm 5$ mas, separation $= 979 \pm 5$ mas, and position angle $= 248.27 \pm 0.29$ deg.

    The target was further observed with the SPHERE/VLT instrument in the IRDIFS mode \citep{2010MNRAS.407...71V,2014A&A...572A..85Z}. More details on these observations, together with the data reduction are presented in \cite{2018arXiv180903848L}. The relative astrometry of the candidate companion with respect to the star were derived from SPHERE/IRDIFS, and the results are shown in Table \ref{tab:astrometry}. The combined astrometry confirms that the companion is bound with the target star. The SPHERE/IFS data was used to derive a low-resolution NIR spectrum \citep{2018arXiv180903848L} which we used to characterise the companion star and estimate its contamination in the \textit{K2} photometry (see section \ref{companion_star}.
    
    \subsection{GAIA Astrometry}
    The \textit{Gaia} Data Release 2 (DR2) has surveyed over one billion stars in the Galaxy \citep{2016A&A...595A...1G,2018arXiv180409365G,2018arXiv180409366L} and provided precise measurements of the parallaxes and proper motions for the sources. K2-265 has a measured parallax of $7.18 \pm 0.05$ mas, corresponding to a distance of $139 \pm 1$ pc. The proper motion of K2-265 is $\mu_{RA} = 30.20 \pm 0.09$ mas, $\mu_{DEC} = -23.34 \pm 0.06$ mas. As part of the \textit{Gaia} DR2, the stellar effective temperature of K2-265 was derived from the three photometric bands \citep{2018arXiv180409374A} as $\rm T_{eff} = 5390_{-53}^{+194}$ K. The G-band extinction $A_g = 0.101$ and the reddening $E(BP-RP) = 0.065$ estimated from the parallax and magnitudes were used to determine the stellar luminosity, which in turn provides an estimate of the stellar radius as $R_s = 0.914_{-0.06}^{+0.02}~R_{\odot}$ \citep{2018arXiv180409374A}. The stellar parameters from the results of \textit{Gaia} DR2 are consistent with the distance estimate, effective temperature and stellar radius which are derived in the joint Bayesian analysis in section \ref{MCMC}. However, \textit{Gaia} DR2 does not detect the companion star in the system and K2-265 is registered as a single object.

\section{Analysis and Results\label{results}}

	\subsection{Spectral Analysis\label{stellarparam}}
	\hspace{0.5cm}The spectral analysis of the host star was performed by co-adding all the individual (Doppler corrected) spectra with IRAF\footnote{IRAF is distributed by National Optical Astronomy Observatories, operated by the Association of Universities for Research in Astronomy, Inc., under contract with the National Science Foundation, USA.}. 
    We first derived the stellar parameters following the analysis of \citet{2008A&A...487..373S} by measuring the equivalent widths (EW) of Fe~{\sc i} and Fe~{\sc ii} lines with version 2 of the ARES code\footnote{The ARES code can be downloaded at http://www.astro.up.pt/~sousasag/ares/} \citep{2015A&A...577A..67S}, and the chemical abundances were derived using the 2014 version of the code MOOG \citep{sneden} which used the iron excitation and ionization balance. We obtained the following parameters: \teff\,=\,5457\,$\pm$\,29\,K, \logg\,=\,4.42\,$\pm$\,0.05\,dex, [Fe/H]\,=\,0.08\,$\pm$\,0.02\,dex, microturbulence $\xi_t$\,=\,0.81\,$\pm$\,0.05\,km\,s$^{-1}$. The errors provided here for the stellar parameters are precision errors which are intrinsic to the method \citep{2011A&A...526A..99S}.
    
    The chemical abundances of the host star are found in Table.~\ref{abundance}. For more details on this analysis and the complete list of lines we refer the reader to the following works:
    \citet{adibekyan12}, \citet{santos15}, and \citet{delgadomena17}. Li and S abundances were derived by spectral synthesis as performed in \citet{delgadomena14} and \citet{ecuvillon04}, respectively.

	\subsection{Characterisation of the Companion Star \label{companion_star}}

To determine the physical parameters of the bound companion, we used the same approach as in \cite{2016ApJ...824...55S}. We fit the magnitude difference between the target and companion star, as observed by SPHERE IRDIFS, with the BT-Settl stellar atmosphere models \citep{2012RSPTA.370.2765A}. The two stars are bound companions (see sections \ref{astrometry}), hence they have the same distance to Earth and age, and they are assumed to have the same iron abundance. We used an MCMC method to derive the companion mass, using the results of the spectral analysis of the target star as priors on the analysis. We used the Dartmouth stellar evolution tracks to convert the companion mass (at a given age and metallicity) into spectroscopic parameters. Our final derivation gives: $\rm T_{eff} = 3428 \pm 22 ~K$, $\rm \log g = 4.870 \pm 0.017 ~[cgs]$, $\rm M_{star B} = 0.40 \pm 0.01 ~M_{\odot}$, $\rm R_{star B} = 0.391_{-0.010}^{+0.006} ~R_{\odot}$, corresponding to a star of spectral type M2 \citep{2000asqu.book.....C}. Using this result, we integrated the SED models in the Kepler band, and derived the contribution of flux contamination in the light curve of star A from star B to be $0.952 \pm 0.024 \%$. The derived contamination of the companion star was taken into account in the joint Bayesian analysis in section \ref{MCMC} to determine the system parameters of K2-265. The parameters of the companion star and their corresponding uncertainties are reported in Table \ref{EPIC1496Param}.

	\subsection{Stellar Rotation\label{K2rotation}}
	Rotational modulation is observed in the detrended \textit{K2} lightcurve as shown in Fig.~\ref{k2lc}. We derived the rotational period of K2-265 using multiple methods to determine the origin of the periodic variation.
	
	We first calculated the stellar rotational period with the auto-correlation-function (ACF) method as described in \citet{2013ApJ...775L..11M,2014ApJS..211...24M}, and found the stellar rotational period as $15.14 \pm 0.38$ d, with a further peak observed at $30.48 \pm 0.28$ d.
	
\begin{figure*}
\centering     
\subfigure[t!][]{\label{periodogramBIS}\includegraphics[width=0.45\textwidth]{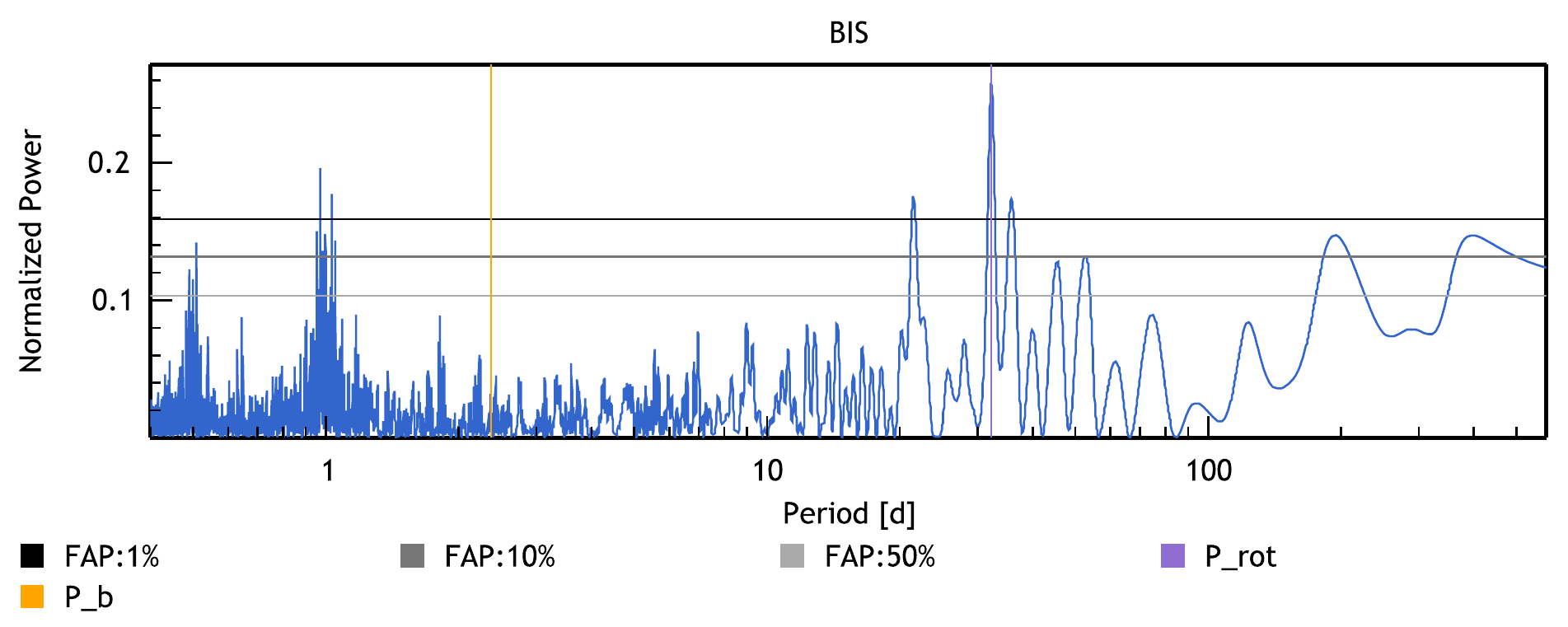}}
\subfigure[t!][]{\label{periodogramFWHM}\includegraphics[width=0.45\textwidth]{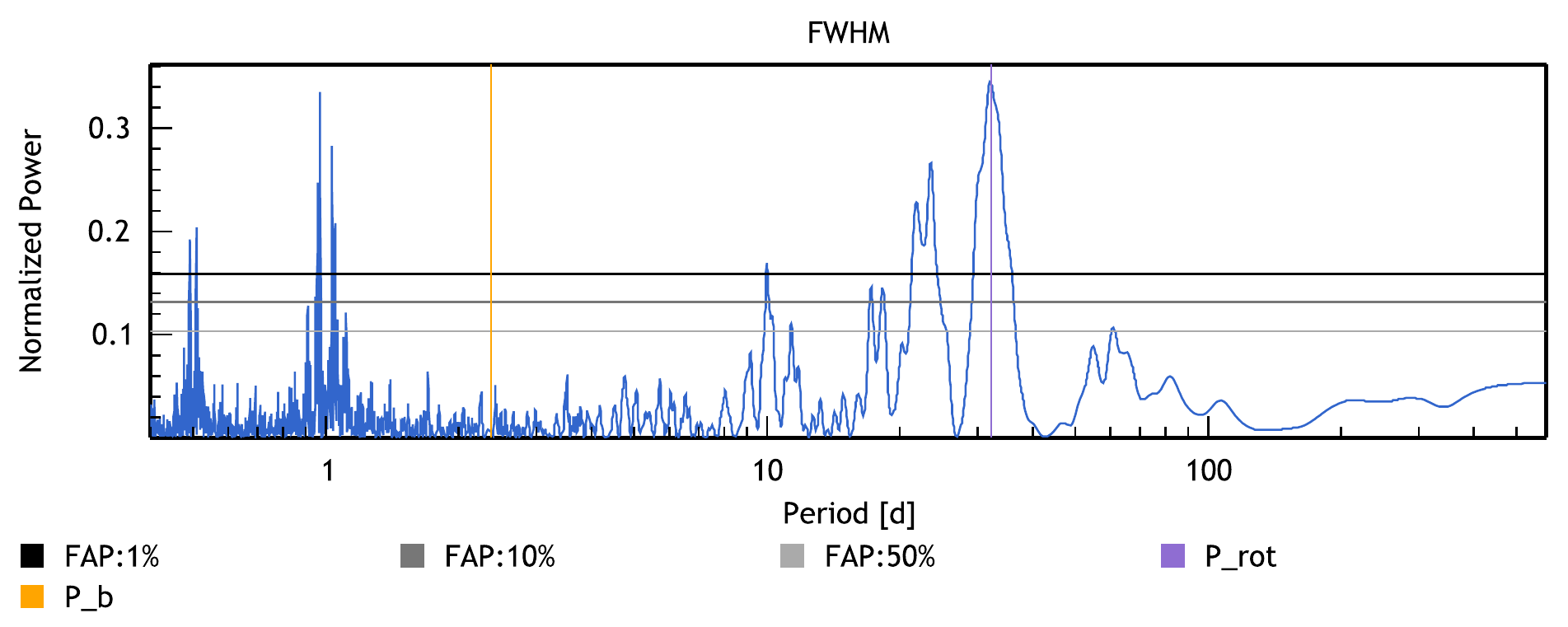}}
\subfigure[t!][]{\label{periodogramRV}\includegraphics[width=0.45\textwidth]{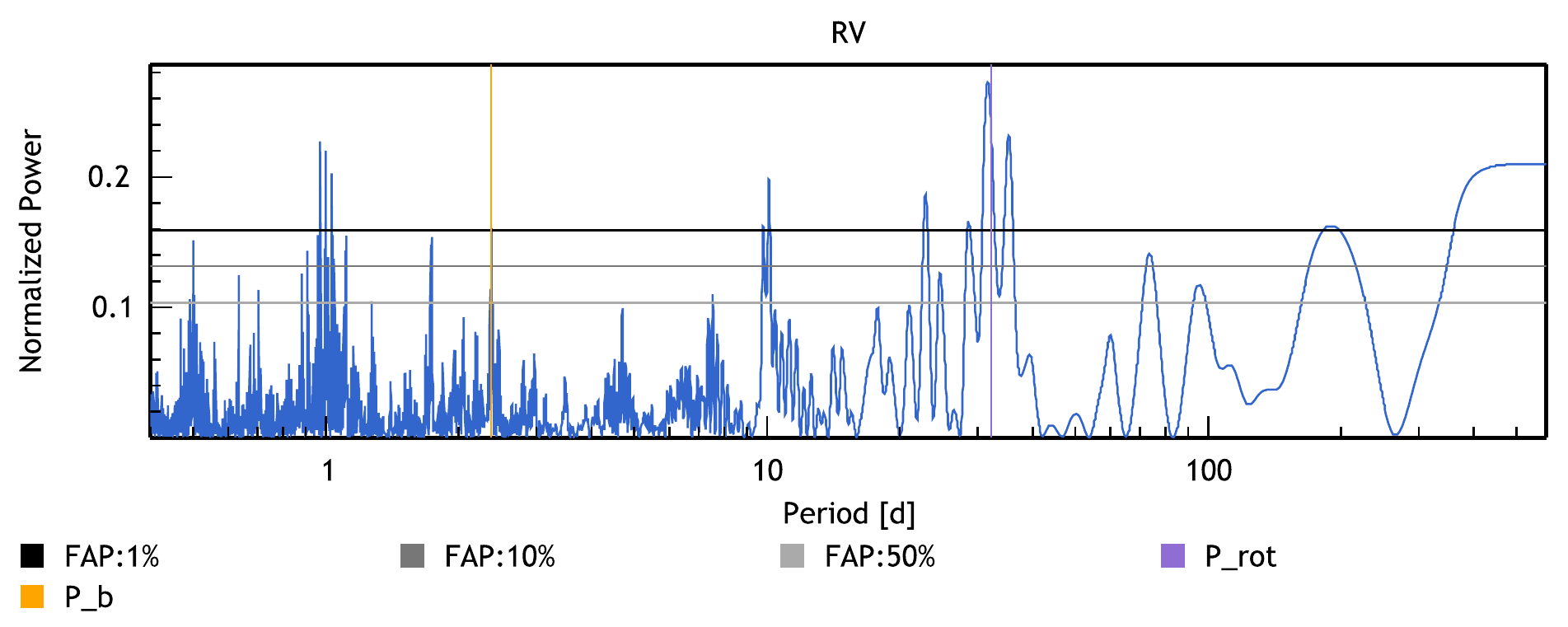}}
\subfigure[t!][]{\label{periodogramSindex}\includegraphics[width=0.45\textwidth]{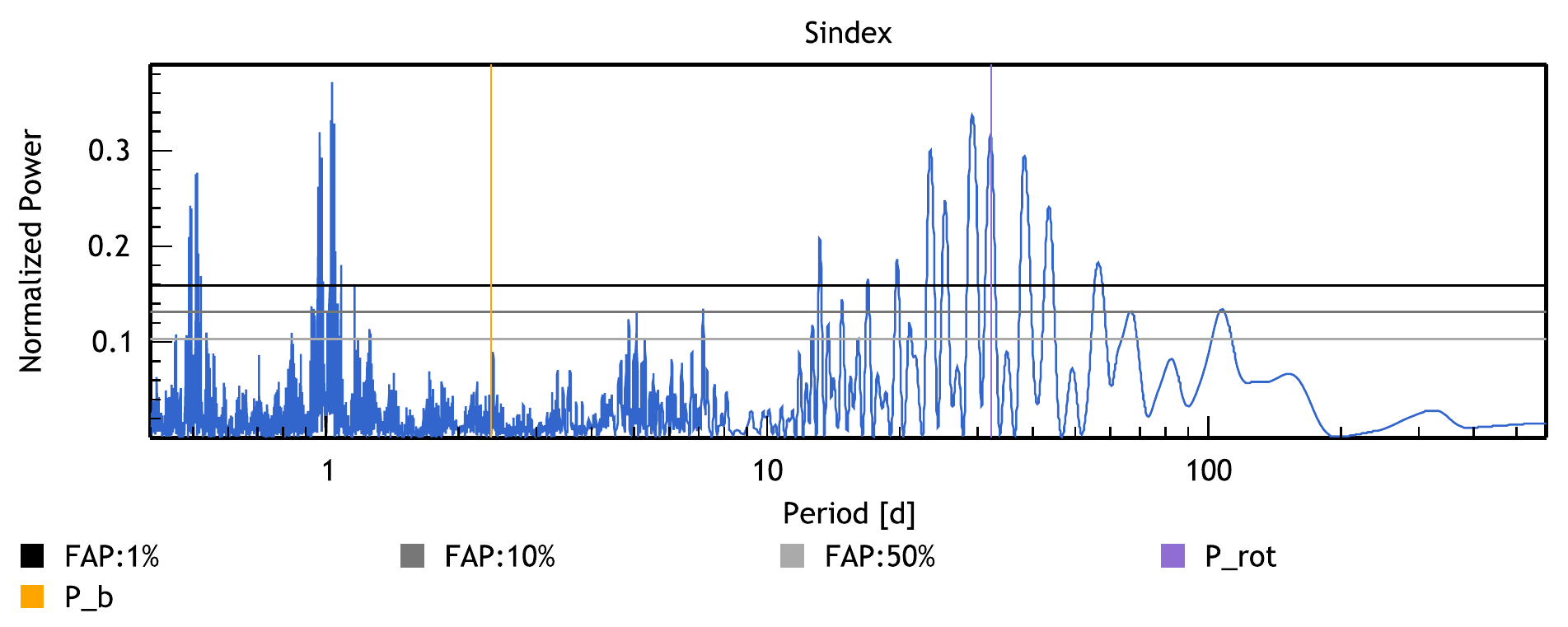}}
\caption{\label{LSplot}Lomb-Scargle periodogram of (a) Bisector Span (BIS); (b) Full Width Half Maximum (FWHM); (c) Radial velocity (RV); (d) S index (S$_{MW}$). The peak position is marked by the purple line and corresponds to a period of $32.2\pm0.6$ d. The orbital period of K2-265 b is indicated by the yellow line, and the planet signal is only significant in the RV at the 1\% FAP level.}
\end{figure*}

	The Lomb-Scargle periodogram \citep{1976Ap&SS..39..447L,1982ApJ...263..835S} analysis was performed to determine the periodicity in the RV data. Fig.~\ref{LSplot} shows the periodogram of the bisector analysis (BIS), the full width at half maximum (FWHM), the RV measurements, and the S index. A clear peak is measured in all four periodograms at $32.2 \pm 0.6$ d, which is larger than but marginally consistent with the ACF period of 30.48 d at a 2-$\sigma$ level. The timescale of lightcurve variation measures the changing visibility of starspots. We attribute the discrepancy between the two rotation periods to latitude variation of the magnetically active regions.
    
    An upper limit of the  sky-projected stellar rotational velocity was derived from the FWHM of the HARPS spectra ($v \sin{i} < 1.9 \pm 0.2$ km s$^{-1}$). Using the stellar radius in Table \ref{EPIC1496Param}, we estimate a rotation period $P_{rot} > 26.02 \pm 3.08$ d (assuming an aligned system, $i = 90^{\circ}$), which agrees with the $\sim30$ d period derived from the photometry and the RV data.

	Furthermore, the stellar rotation period was also derived following the method of \cite{2008ApJ...687.1264M}. In summary, we used the $\rm B-V$ colour from APASS\footnote{https://www.aavso.org/apass} to find the convective turnover time $\tau_c$ using calibrations from \cite{1984ApJ...287..769N}. We then used the measured Mount Wilson index $S_{MW} = 0.195 \pm 0.025$ to derive $\rm \log R'_{HK} = -4.90 \pm 0.12$, from which we determine the Rossby number $\rm R_o = 1.94$ using calibrations from \cite{2008ApJ...687.1264M}. Finally, using the relation $\rm P_{rot} = R_o \ \times$ $\tau_{c}$, we calculated the stellar rotation period as $32 \pm 10$ d.

	\subsection{Joint Bayesian Analysis With \texttt{PASTIS} \label{MCMC}}
	We employed a Bayesian approach to derive the physical parameters of the host star and the planet. We jointly analysed the \textit{K2} photometric light curve, the HARPS RV measurements and the spectral energy distribution (SED) observed by the APASS, 2-MASS, and WISE surveys (\citealt{2014AJ....148...81M,2014yCat.2328....0C}; a full list of host star magnitudes can be found in Table \ref{PhotometricProperties}) using the \texttt{PASTIS} software \citep{2014MNRAS.441..983D,2015MNRAS.451.2337S}. The light curve was modelled using the \texttt{jktebop} package \citep{2008MNRAS.386.1644S} by taking an oversampling factor of 30 to account for the long integration time of the K2 data \citep{2010MNRAS.408.1758K}. The RVs were modelled with Keplerian orbits. Following similar approaches to \cite{2017A&A...608A..25B} and \cite{2018NatAs...2..393S}, a Gaussian process (GP) regression was used to model the activity signal of the star. The SED was modelled using the BT-Settl library of stellar atmosphere models \citep{2012RSPTA.370.2765A}. \newline  
	
	The Markov Chain Monte Carlo (MCMC) method was used to derive the system parameters. The spectroscopic parameters of K2-265A were converted into physical stellar parameters using the Dartmouth evolution tracks \citep{2008ApJS..178...89D} at each step of the chain. The quadratic limb darkening coefficients were also computed using the stellar parameters and tables of \cite{2011A&A...529A..75C}. 
	
	For the stellar parameters, we used normal distribution priors centred on the values derived in our spectral analysis. We chose a normal prior for the orbital ephemeris centred on values found by the detection pipeline. Furthermore, we adopted a sine distribution for the inclination of the planet. Uninformative priors were used for the other parameters. The priors of the fitted parameters used in the model can be found in Table \ref{MCMCprior}.
	
	Twenty MCMC chains of $3 \times 10^5$ iterations were run during the MCMC analysis, where the starting points were randomly drawn from the joint prior. The Kolmogorov-Smirnov test was used to test for convergence in each chain. We then removed the burn-in phase and merged the converged chains to derive the system parameters.

	\begin{table}[!t]
		\caption{\label{EPIC1496Param}System parameters of K2-265 obtained from \texttt{PASTIS}. Stellar parameters of Star B were derived as described in section \ref{companion_star}.}
		\small
		\centering
		\begin{tabular}{ l c}
		\hline 
		Parameter & Value and uncertainty \\ 
		\hline
        \textit{Stellar parameters} \\
        \textit{Star A} \\
 		Effective temperature \teff\ [K] 	&	$5477 \pm 27$ \\
		Surface gravity \logg\ [cgs] & $4.419\pm0.053$ \\
		Iron abundance \met\ [dex] & $0.078\pm0.020$ \\
		Distance to Earth $D$ [pc]	& $145\pm8$\\
		Interstellar extinction $E(B-V)$ [mag]	& $0.009^{_{+0.011}}_{^{-0.007}}$ \\
		Systemic radial velocity $\gamma$ [\kms] & $-18.186\pm0.002$ \\
		Stellar density $\rho_{\star}/\rho_{\astrosun}$	& $0.98\pm0.19$	\\
		Stellar mass M$_{\star}$\ [\Msun] & $0.915\pm0.017$ \\
		Stellar radius R$_{\star}$\ [\Rsun] &	$0.977\pm0.053$ \\
		Stellar age $\tau$\ [Gyr] &	$9.7\pm3.0$	\\
		&\\
        
        \textit{Star B}\\
        Effective temperature \teff\ [K] 	&	$3428 \pm 22$ \\
		Surface gravity \logg\ [cgs] & $4.870\pm0.017$ \\
		Stellar mass M$_{\star}$\ [\Msun] & $0.40\pm0.0.01$ \\
		Stellar radius R$_{\star}$\ [\Rsun] &	$0.391_{-0.010}^{0.006}$ \\
        
        &\\
        
		\hline
		\textit{Planet Parameters}\\
		Orbital Period $P$ [d] &	$2.369172\pm8.9\times10^{-5}$ \\
		Transit epoch $T_{0}$ [BJD - 2456000] 	& $981.6431\pm1.6\times10^{-3}$	\\
		Radial velocity semi-amplitude $K$ [\ms] & $3.34\pm0.43$	\\
		Orbital inclination $i$ [$^{\circ}$] 	& $87.7\pm1.6$ \\
		Planet-to-star radius ratio $k$ & $0.01604\pm0.00041$	\\
		Orbital eccentricity $e$	& $0.084\pm0.079$	\\
		Impact parameter $b$ & $0.30\pm0.20$	\\
		Transit duration T$_{14}$ [h] & $2.266\pm0.050$	\\
		Semi-major axis $a$ [AU] & $0.03376\pm0.00021$ 	\\
		Planet mass M$_{p}$ [\Mearth]	& $6.54 \pm 0.84$	\\
		Planet radius R$_{p}$ [\Rearth] &	$1.71\pm0.11$		\\
		Planet bulk density $\rho_{p}$ [\gcm3] & $7.1\pm1.8$\\
		&\\
		\hline
		\end{tabular} 
		\end{table}

	\subsection{Stellar Age\label{age}}
	\hspace{0.5cm}From the joint analysis of the observational data, together with the Dartmouth stellar evolution tracks, the age of K2-265 was determined as $\tau_{iso} = 9.7\pm3.0$ Gyr. The stellar rotation analysis in Section \ref{K2rotation} found that K2-265 has a rotation period of $\sim 30$ d. We adopted a rotational period of $32.2 \pm 0.6$ d, and followed the methods by \cite{2010ApJ...722..222B} to find that K2-265 has a gyrochronological age of $\tau_{gyro} = 5.34 \pm 0.19$ Gyr. We further derived the age of K2-265 using the relation between the [Y/Mg] abundance ratio and the stellar age \citep{2016A&A...590A..32T,2015A&A...579A..52N}, and found an age of $\tau_{[Y/Mg]} = 3.97 \pm 2.59$ Gyr. $\tau_{[Y/Mg]}$ agrees with $\tau_{gyro}$ within 1-$\sigma$ uncertainty but is lower than the derived isochronal age. The low lithium abundance A(Li~{\sc ii})$ < 0.45$ of the host star obtained from spectral analysis (Section \ref{stellarparam}) suggests that the host is not young. Hence it is likely that the host is of at least an intermediate age.

\section{Discussion \& Conclusion\label{conclusion}}
\hspace{0.5cm}K2-265 b has a mass of $\rm 6.54 \pm 0.84~M_{\oplus}$ and a radius of $\rm 1.71 \pm 0.11~R_{\oplus}$. This corresponds to a bulk density of $7.1 \pm 1.8~ \rm g\ cm^{-3}$, which is slightly higher than that of the Earth's density. We applied a number of theoretical models to investigate the planet's interior composition. 

\citet{2007ApJ...659.1661F} modelled the radii of planets with a range of different masses at various compositions, and derived an analytical function which allows an estimate of the rock mass fraction (rmf) of ice-rock-iron planets. We find a rmf of 0.84 for K2-265 b which is equivalent to a rock-to-iron ratio of $0.84/0.16$, a rock fraction that is higher than the Earth. \citet{2007ApJ...669.1279S} also used interior models of planets to study the mass-radius relation of solid planets. By assuming the planets are composed primarily of iron, silicates, water, and carbon compounds, \citet{2007ApJ...669.1279S} showed that masses and radii of terrestrial planets follow a power law. Using the derived best-fit mass and radius of K2-265 b, the bulk composition of the planet was determined to be predominantly rocky with $>70\%$ of silicate mantle by mass.

		\begin{figure}[htbp!]
	    \includegraphics[width=0.5\textwidth]{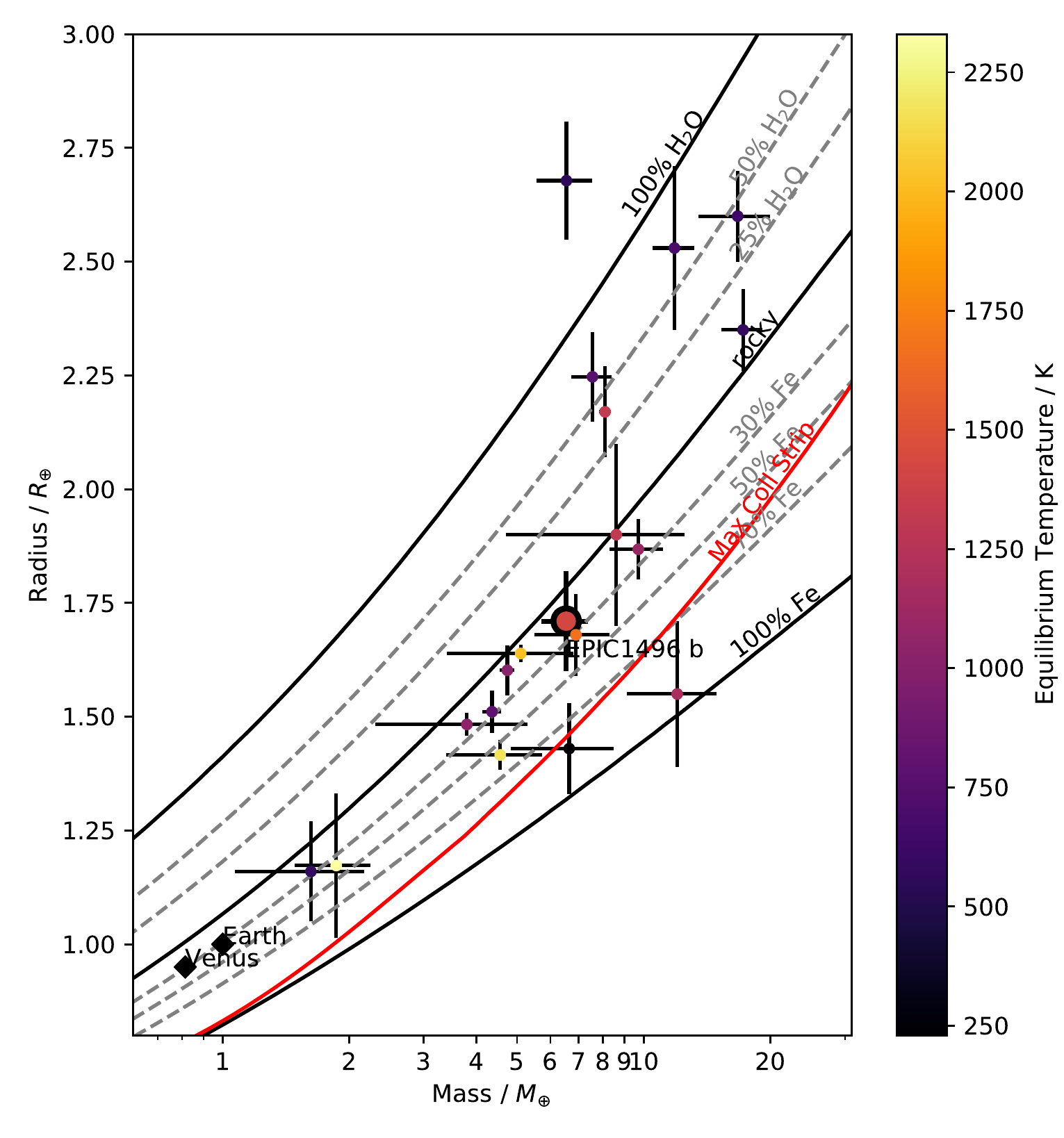}
	    \caption{\label{MRplot} A mass-radius diagram of confirmed Earth-sized planets with masses up to $20\ \rm M_{\oplus}$. Data were taken from NASA Exoplanet archive \protect\footnotemark . The mass-radius relations were taken from \cite{2016ApJ...819..127Z}. From top to bottom, the black solid lines denotes a pure water, pure rock and pure iron composition. The grey dashed lines between the solid lines are mass-radius relations for water-rock and rock-iron composites. The red solid line is the lower limit of a planet radius as a result of collisional stripping \citep{2010ApJ...712L..73M}. K2-265 b has a composition consistent with a rocky terrestrial planet.	}
	\end{figure}	
	
\footnotetext{\url{https://exoplanetarchive.ipac.caltech.edu/index.html}}

\begin{figure}[htbp!]
	    \includegraphics[width=0.5\textwidth]{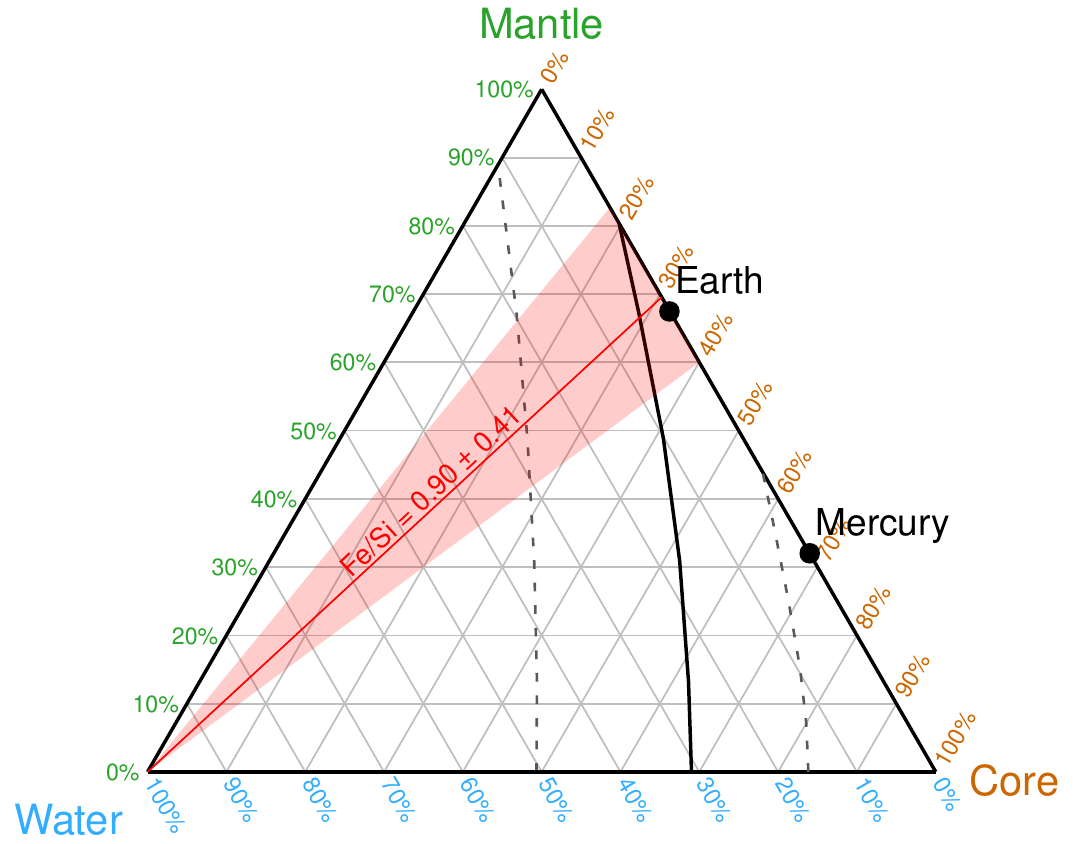}
	    \caption{\label{ternary}Ternary diagram showing the possible composition of K2-265 b. The thick black line is the allowed composition of the planet inferred from the central values of the planet’s mass and radius, and the dashed lines denote the deviations from this line allowed by the 1-$\sigma$ uncertainties on the fundamental parameters. The red line and area show the compositions compatible with the planetary bulk Fe/Si ratio derived for K2-265 b from the stellar value. Compositions of the Earth and Mercury are shown for reference.}.
	\end{figure}

We performed a more detailed investigation of the composition of K2-265 b using the interior model of \cite{2017ApJ...850...93B}. This model considers planets made out of three differentiated layers: core (metals), mantle (rocks), and a liquid water envelope. Figure \ref{ternary} shows the possible compositions of K2-265 b inferred from the 1-$\sigma$ uncertainties on the planet’s mass and radius. By focusing on terrestrial compositions only (i.e. without any water), we show that the central mass and radius of the planet are best fitted with a rock mass fraction of 81\%, consistently with other theoretical predictions. However, given the uncertainties on the fundamental parameters, the rmf remains poorly constrained, namely within the 44–100\% range. If we assume that the stellar Fe/Si ratio (here $0.90\pm0.41$) can be used as a proxy for the bulk planetary value \citep{2015A&A...577A..83D,2017ApJ...850...93B}, this range is reduced to 60–83\%. In the case of a water-rich K2-265 b, the model only allows us to derive an upper limit on the planet’s water mass fraction (wmf). Indeed, given the high equilibrium temperature of the planet ($\sim1300$ K assuming an Earth-like albedo), water would be in the gaseous and supercritical phases, which are less dense than the liquid phase. From the uncertainties on the mass, radius, and bulk Fe/Si ratio of K2-265 b, we infer that this planet cannot present a wmf larger than 31\%.

The California-\textit{Kepler} Survey (CKS) measured precise stellar parameters of Kepler host stars using spectroscopic follow-up \citep{2017AJ....154..108J}, and refined the planetary radii to study the planet size distribution and planet occurrence rate \citep{2017AJ....154..109F}. The survey has revealed a bimodal distribution of small planet sizes. Planets tend to have radii of either $\approx1.3 \rm ~R_{\oplus}$ or $\approx2.4 \rm ~R_{\oplus}$, with a deficit of planets at $\approx1.8 \rm ~R_{\oplus}$. The survey confirms the prediction by \citet{2013ApJ...775..105O}, whereby a gap in the planetary radius distribution exists as a consequence of atmospheric erosion by the photoevaporation mechanism. Alternatively, the core-powered mass loss mechanism could also drive the evaporation of small planets \citep{2016ApJ...825...29G, 2018MNRAS.476..759G}.

Due to its close proximity to the host star, the super-Earth K2-265 b is exposed to strong stellar irradiation. The planet’s gas envelope could be evaporated as a result. This process was observed in a number of systems (e.g. HD209458 b ; \citealt{2003Natur.422..143V}, GJ 436 b; \citealt{2015Natur.522..459E}). The present irradiance of the planet is $\rm S = S_{\oplus}(L_s/L_{\odot})(AU/a)^2 = 9.32 \times 10^5~W m^{-2}$, where $\rm L_s$ and $\rm L_{\odot}$ are the luminosity of the star and the Sun, $\rm S_{\oplus}$ is the Solar irradiance on Earth, and $\rm a$ is the semi-major axis of the planet. The equilibrium temperature of the K2-265 b can be estimated using Equ.~1 of \cite{2007ApJ...667L.191L}: $\rm T_{eq} = T_{eff}(R_s/a)^{1/2}[f(1-A_B)]$, where f and $\rm A_B$ are the reradiation factor and the Bond albedo of the planet. Assuming an Earth-like Bond albedo $\rm A_B=0.3$ and that the incident radiation is redistributed around the atmosphere (i.e. $\rm f=1/4$), the equilibrium temperature of K2-265 b is $\rm T_{eff} \approx 1300$ K.

\begin{figure}[htbp!]
        \includegraphics[width=0.5\textwidth]{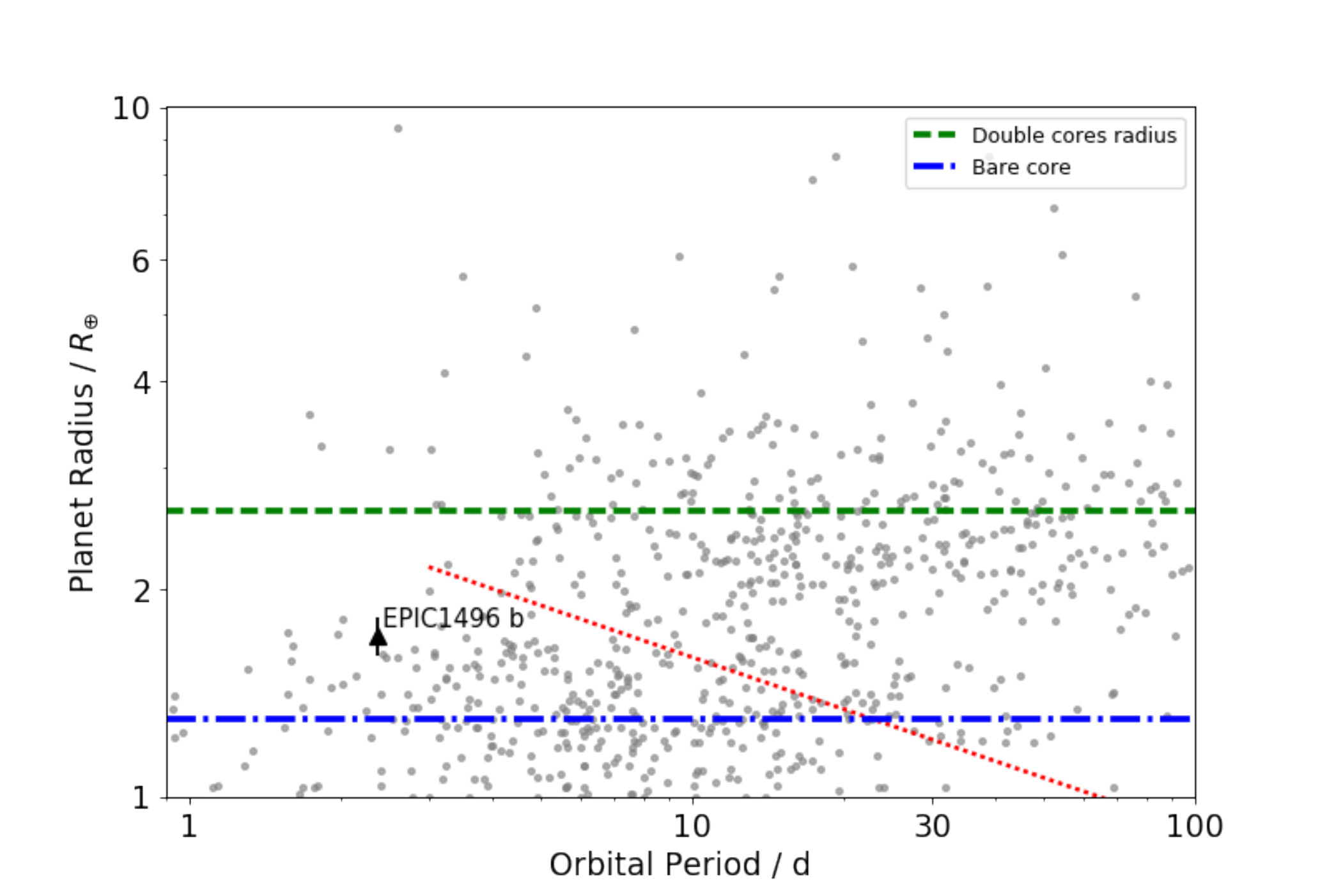}
        \caption{\label{evaporationvalley} Planet radius distribution as a function of orbital period. The grey circles denote the planet sample obtained from the CKS sample \citep{2017AJ....154..109F}. The blue dot-dashed line and the green dashed line indicate the peak of the bimodal distribution of the planet radius distribution, where planets tend to favour radii of $\sim 1.3~R_{\oplus}$ and $\sim 2.4~R_{\oplus}$ due to the photoevaporation mechanism. The red dotted line indicates the lower limit of the photoevaporation valley derived from \cite{2017ApJ...847...29O}.}
        
    \end{figure}

Indeed, K2-265 b lies below the lower limit of the photoevaporation valley as shown in the 2D radius distribution plot in Figure \ref{evaporationvalley}. This implies that the planet could have been stripped bare due to photoevaporation, revealing its naked core. This atmospheric stripping process is presumed to have occurred in the first $\approx100$ Myr since the birth of the planet when X-ray emission is saturated \citep{2012MNRAS.422.2024J}, after which the X-ray emission decays. We estimated the total X-ray luminosity of K2-265 over its lifetime, $E_x^{tot}$, using the X-ray-age relation of \cite{2012MNRAS.422.2024J}. Using the results of section \ref{age}, we adopted a mean age of 6.32 Gyr for the host star. The X-ray-to-bolometric luminosity ratio in the saturated regime for a $B-V = 0.743$ star is $\log{(L_x/L_{bol})} = -3.71 \pm 0.05 \pm0.47$. The corresponding turn-off age is $\log{\tau_{sat}} = 8.03 \pm 0.06 \pm 0.31$, where the decrease in X-ray emission follows a power law ($\alpha = 1.28 \pm 0.17$). Over the lifetime of the star, $E_x^{tot} = 6.70 \times 10^{45}$ ergs (assuming efficiency factor $\eta = 0.25$) and K2-265 b is expected to have lost 2.7\% of its mass under the constant-density assumption. K2-265 b has a predominantly rocky interior as shown in Figure \ref{ternary}. This indicates that the planet was likely formed inside the ice-line, and could have either migrated to its current orbital separation well before $\approx100$ Myr or accreted its mass locally \citep{2017ApJ...847...29O}.

K2-265 b is among the denser super-Earths below the photoevaporation gap. In addition to photoevaporation, giant impact between super-Earths could drive mass loss in the planetary atmosphere. Super-Earths are thought to have formed via accretion in gas discs, followed by migration and eccentricity damping due to their interactions with the gas disc (e.g. \citealt{2015ApJ...811...41L}), leading to densely packed planetary systems. As the gas disc disperses, secular perturbation between planets excites their eccentricity, triggering giant impacts between the bodies before the system becomes stable \citep{2014A&A...569A..56C}. Two planets of comparable sizes could collide at a velocity beyond the surface escape velocity \citep{2004ApJ...613L.157A,2009ApJ...700L.118M}. The impact could lead to a reduction in the planet envelope-to-core-mass ratio, hence an increase in the mean density and alteration of the bulk composition of the planet \citep{2015ApJ...812..164L, 2016ApJ...817L..13I}.

Discoveries of super-Earths have shown a diversity of small planets in the mass-radius diagram. Precise RV and photometric measurements with an accuracy of a few percent are necessary to put strong constraints on the planetary mass and radius, and provide a precise bulk composition. The core composition of the planet can be derived as a result. In particular, the mass fraction of a planetary core can inform us of the formation and evolution history of the planet. K2-265 b has a precisely determined mass (13\%) and radius (6\%), and the composition of the planet is consistent with a rocky planet. Its small radius and short orbital period suggest that K2-265 b could have been photoevaporated to a bare rocky core. Its high rock-to-mass fraction implies a planet formation within the ice line. Studying planets with an exposed core could provide valuable insight to planet formation via the core accretion mechanism. The increasing sample of small planets will help distinguish planet origins, identify types of mass loss mechanism, and probe the efficiency of atmospheric evaporation processes.

\begin{acknowledgements}
We thank the anonymous referee for the helpful comments which improved the manuscript. 
DJAB acknowledges support by the UK Space Agency. DJA gratefully acknowledges support from the STFC via an Ernest Rutherford Fellowship (ST/R00384X/1).
This work was funded by FEDER - Fundo Europeu de Desenvolvimento Regional funds through the COMPETE 2020 - Programa Operacional Competitividade e Internacionalização (POCI), and by Portuguese funds through FCT - Fundação para a Ciência e a Tecnologia in the framework of the projects POCI-01-0145-FEDER-028953 and POCI-01-0145-FEDER-032113.
N.S. and O.D. also acknowledge the support from FCT and FEDER through COMPETE2020 to grants UID/FIS/04434/2013 \& POCI-01-0145-FEDER-007672, PTDC/FIS-AST/1526/2014 \& POCI-01-0145-FEDER-016886 and PTDC/FIS-AST/7073/2014\& POCI-01-0145-FEDER-016880. 
S.G.S acknowledge support from FCT through Investigador FCT contract nr. IF/00028/2014/CP1215/CT0002.
SCCB also acknowledges support from FCT through Investigador FCT contracts IF/01312/2014/CP1215/CT0004.
E.D.M. acknowledges the support by the Investigador FCT contract IF/00849/2015/CP1273/CT0003.
RL has received funding from the European Union’s Horizon 2020 research and innovation programme under the Marie Skłodowska-Curie grant agreement n.664931.
FF acknowledges support from  PLATO ASI-INAF contract n.2015-019-R0.
This research was made possible through the use of the AAVSO Photometric All-Sky Survey (APASS), funded by the Robert Martin Ayers Sciences Fund.  This publication makes use of data products from the Two Micron All Sky Survey, which is a joint project of the University of Massachusetts and the Infrared Processing and Analysis Center/California Institute of Technology, funded by the National Aeronautics and Space Administration and the National Science Foundation. This publication makes use of data products from the Wide-field Infrared Survey Explorer, which is a joint project of the University of California, Los Angeles, and the Jet Propulsion Laboratory/California Institute of Technology, funded by the National Aeronautics and Space Administration. This research has made use of NASA's Astrophysics Data System Bibliographic Services.
This paper includes data collected by the K2 mission. Funding for the K2 mission is provided by the NASA Science Mission directorate.
This research has made use of the Exoplanet Follow-up Observation Program website, which is operated by the California Institute of Technology, under contract with the National Aeronautics and Space Administration under the Exoplanet Exploration Program.
This work has made use of data from the European Space Agency (ESA) mission
{\it Gaia} (\url{https://www.cosmos.esa.int/gaia}), processed by the {\it Gaia}
Data Processing and Analysis Consortium (DPAC,
\url{https://www.cosmos.esa.int/web/gaia/dpac/consortium}). Funding for the DPAC
has been provided by national institutions, in particular the institutions
participating in the {\it Gaia} Multilateral Agreement.

\end{acknowledgements}

\bibliographystyle{aa}
\bibliography{EPIC1496}{}

\begin{appendix}
\section{Supplementary Tables and Figures}


\begin{figure}[h]
\begin{center}
\includegraphics[width=\figw]{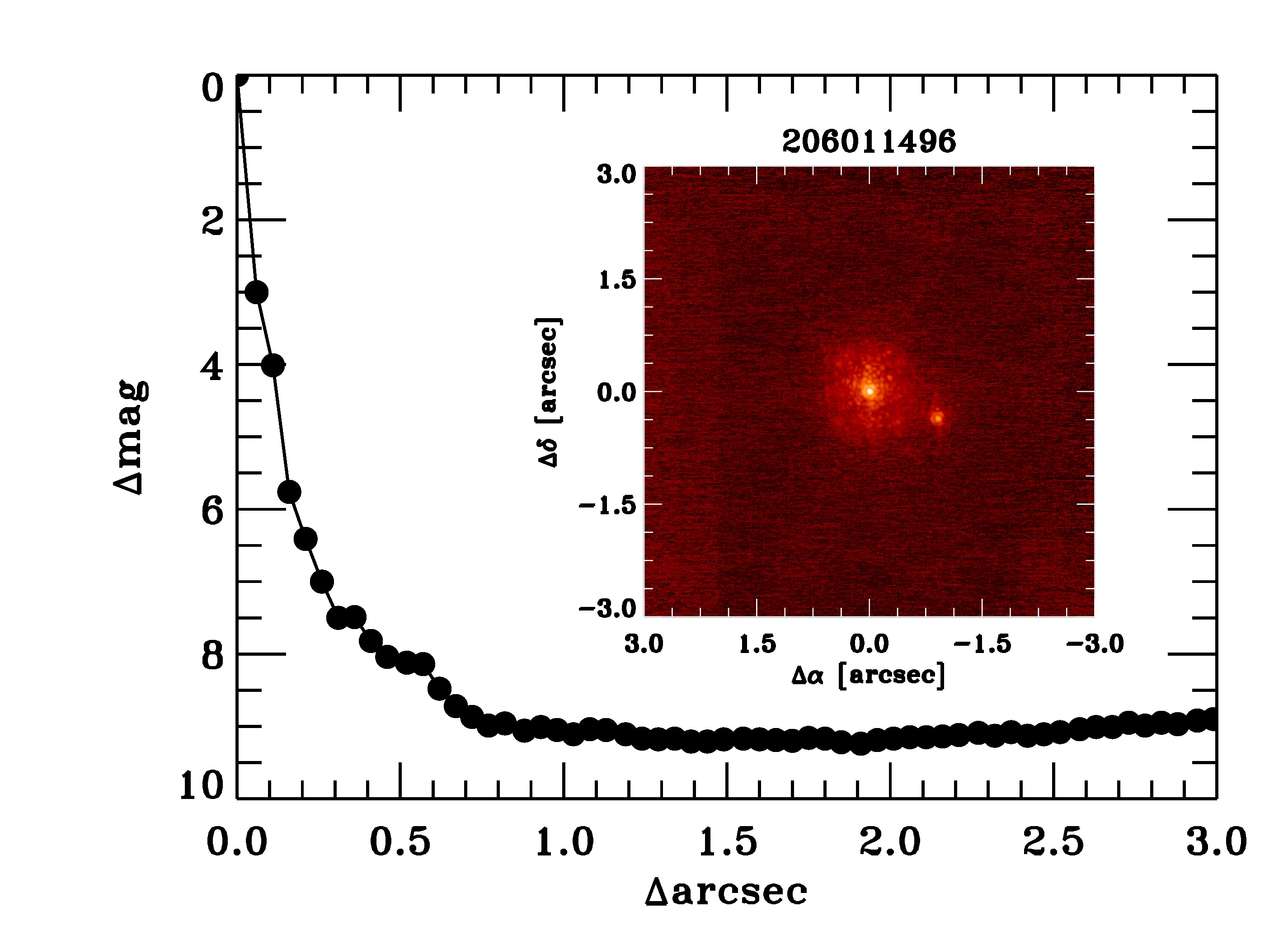}
\label{tbd}
\caption{\label{KECKAO}K-band Keck AO image shows a companion at a separation of $= 979 \pm 5$ mas.}
\end{center}
\end{figure}

\begin{table*}
  \caption{NIRC2 and IRDIS astrometry of the candidate companion}
  \label{tab:astrometry}
  \centering
  \begin{tabular}{cccccc}
    \hline\hline
    Instrument & Date       & $\Delta\alpha$ & $\Delta\delta$ & Sep.        & Pos. ang. \\
               &            & (mas)          & (mas)          & (mas)       & (deg)     \\
    \hline
    Keck/NIRC2 & 2015-08-04 & -910 $\pm$ 5   & -363 $\pm$ 5   & 979 $\pm$ 5 & 248.27 $\pm$ 0.29 \\
    VLT/SPHERE & 2015-08-04 & -906 $\pm$ 1   & -368 $\pm$ 1   & 978 $\pm$ 1 & 247.87 $\pm$ 0.20 \\
    VLT/SPHERE & 2017-08-30 & -903 $\pm$ 1   & -365 $\pm$ 1   & 975 $\pm$ 1 & 247.99 $\pm$ 0.01 \\
    \hline
  \end{tabular}
\end{table*}

\begin{table*}
\centering
\caption{\label{abundance}Chemical abundances of the host star, relative to the Sun.}
\medskip
\begin{tabular}{lcc}
\hline
Element & Abundance& Lines number\\
 $[$X/H$]$  & [dex] & \\
\hline
\ion{C}{1} & $0.01 \pm 0.05$ & 2\\
\ion{O}{1} & $0.14 \pm 0.10$ & 2\\
\ion{Na}{1} & $0.059 \pm 0.023$ & 2\\
\ion{Mg}{1} & $0.068 \pm 0.068$ & 3\\
\ion{Al}{1} & $0.012 \pm 0.023$ & 2\\
\ion{Si}{1} & $0.053 \pm 0.037$ & 11\\
\ion{S}{1} & $0.05 \pm 0.08$ & 2\\
\ion{Ca}{1} & $0.102 \pm 0.051$ & 9\\
\ion{Sc}{1} & $0.081 \pm 0.053$ & 3\\
\ion{Sc}{2} & $0.099 \pm 0.026$ & 6\\
\ion{Ti}{1} & $0.117 \pm 0.045$ & 18\\
\ion{Ti}{2} & $0.064 \pm 0.034$ & 5\\
\ion{V}{1} & $0.186 \pm 0.052$ & 6\\
\ion{Cr}{1} & $0.088 \pm 0.036$ & 17\\
\ion{Mn}{1} & $0.128 \pm 0.049$ & 5\\
\ion{Co}{1} & $0.130 \pm 0.04$ & 7\\
\ion{Ni}{1} & $0.069 \pm 0.023$ & 40\\
\ion{Cu}{1} & $0.10 \pm 0.04$ & 4\\
\ion{Zn}{1} & $0.00 \pm 0.02$ & 3\\
\ion{Sr}{1} & $0.17 \pm 0.08$ & 1\\
\ion{Y}{2} & $0.09 \pm 0.04$ & 6\\
\ion{Zr}{2} & $0.13 \pm 0.04$ & 4\\
\ion{Ba}{2} & $0.07 \pm 0.04$ & 3\\
\ion{Ce}{2} & $0.13 \pm 0.07$ & 4\\
\ion{Nd}{2} & $0.11 \pm 0.03$ & 2\\
\hline
A(\ion{Li}{1})$^{\ast}$ & $< 0.45$ & 1\\
\hline
\multicolumn{3}{l}{$^{\ast}$A(Li) = log[N(Li)/N(H)] + 12}
\end{tabular}
\end{table*}

\newpage
\onecolumn

\begin{longtable}{cccccccccc}
\caption{\label{RVtable}\textbf{Radial velocity data.} The Barycentric Julian Date (BJD) is given with an offset of 2400000. Signal-to-noise ratio (S/N) is given per CCD pixel at 550nm.}\\
\hline
Time & RV & $\sigma$RV & FWHM & $\sigma$FWHM & BIS & $\sigma$BIS & S$_{\rm MW}$ & $\sigma$S$_{\rm MW}$ & S/N\\
$[$BJD$]$ & [\kms] & [\ms] & [\kms] & [\ms] & [\ms] & [\ms]  & & &  \\
\hline
\endfirsthead
\multicolumn{10}{l}{{\bfseries \tablename\ \thetable{} -- continued from previous page}} \\
\hline
Time & RV & $\sigma$RV & FWHM & $\sigma$FWHM & BIS & $\sigma$BIS & S_{\rm MW} & $\sigma$S_{\rm MW} & S/N\\
\hline
\endhead
\multicolumn{10}{l}{{Continued on next page}} \\ 
\hline
\endfoot
\hline
\endlastfoot
57690.54527  &  -18.19493  &  2.01  &  6.9428  &  4.0  &  -27.1  &  4.0  &  0.1820  &  0.0064  &  46.8  \\
57690.65420  &  -18.19306  &  1.70  &  6.9393  &  3.4  &  -24.5  &  3.4  &  0.1826  &  0.0056  &  57.3  \\
57691.52542  &  -18.18932  &  1.84  &  6.9423  &  3.7  &  -27.3  &  3.7  &  0.1755  &  0.0056  &  51.0  \\
57691.64089  &  -18.18784  &  1.57  &  6.9380  &  3.1  &  -19.0  &  3.1  &  0.1789  &  0.0052  &  62.8  \\
57692.54337  &  -18.19043  &  1.82  &  6.9336  &  3.6  &  -17.0  &  3.6  &  0.1799  &  0.0057  &  52.2  \\
57692.66890  &  -18.19007  &  2.03  &  6.9400  &  4.1  &  -18.4  &  4.1  &  0.1868  &  0.0082  &  48.1  \\
57694.55555  &  -18.18570  &  1.90  &  6.9404  &  3.8  &  -30.9  &  3.8  &  0.1873  &  0.0061  &  50.0  \\
57694.65367  &  -18.18799  &  1.89  &  6.9472  &  3.8  &  -17.4  &  3.8  &  0.1825  &  0.0071  &  51.6  \\
57695.53842  &  -18.19169  &  2.29  &  6.9375  &  4.6  &  -34.0  &  4.6  &  0.1877  &  0.0077  &  40.8  \\
57695.55993  &  -18.19412  &  2.09  &  6.9479  &  4.2  &  -24.2  &  4.2  &  0.1730  &  0.0065  &  44.5  \\
57696.54203  &  -18.19052  &  2.05  &  6.9560  &  4.1  &  -26.0  &  4.1  &  0.1857  &  0.0068  &  45.8  \\
57696.67629  &  -18.19411  &  1.94  &  6.9507  &  3.9  &  -22.7  &  3.9  &  0.1817  &  0.0075  &  50.1  \\
57697.56269  &  -18.19541  &  1.90  &  6.9536  &  3.8  &  -29.4  &  3.8  &  0.1788  &  0.0061  &  49.7  \\
57697.64405  &  -18.19883  &  1.74  &  6.9470  &  3.5  &  -26.9  &  3.5  &  0.1855  &  0.0058  &  55.5  \\
57699.51415  &  -18.18805  &  1.53  &  6.9476  &  3.1  &  -24.6  &  3.1  &  0.2188  &  0.0041  &  62.7  \\
57699.56041  &  -18.18795  &  1.56  &  6.9584  &  3.1  &  -18.3  &  3.1  &  0.2205  &  0.0048  &  62.6  \\
57701.53183  &  -18.18327  &  2.03  &  6.9532  &  4.1  &  -26.0  &  4.1  &  0.2197  &  0.0061  &  45.9  \\
57701.57985  &  -18.18358  &  2.01  &  6.9494  &  4.0  &  -26.0  &  4.0  &  0.2343  &  0.0061  &  46.5  \\
57703.53660  &  -18.18968  &  1.74  &  6.9375  &  3.5  &  -15.2  &  3.5  &  0.2164  &  0.0054  &  55.2  \\
57703.57238  &  -18.18313  &  1.78  &  6.9565  &  3.6  &  -15.9  &  3.6  &  0.2190  &  0.0055  &  53.8  \\
57705.53138  &  -18.19312  &  1.53  &  6.9390  &  3.1  &  -12.6  &  3.1  &  0.2031  &  0.0045  &  64.9  \\
57705.57766  &  -18.19513  &  1.76  &  6.9337  &  3.5  &  -16.9  &  3.5  &  0.2142  &  0.0060  &  55.6  \\
57714.60057  &  -18.19074  &  2.33  &  6.9600  &  4.7  &  -13.1  &  4.7  &  0.2277  &  0.0086  &  41.2  \\
57714.62010  &  -18.19333  &  2.37  &  6.9489  &  4.7  &  -12.6  &  4.7  &  0.2225  &  0.0099  &  41.4  \\
57717.55993  &  -18.19042  &  2.01  &  6.9418  &  4.0  &  -17.8  &  4.0  &  0.1746  &  0.0073  &  48.1  \\
57717.58112  &  -18.18483  &  2.13  &  6.9487  &  4.3  &  -17.2  &  4.3  &  0.1940  &  0.0079  &  45.2  \\
57718.53008  &  -18.19030  &  1.78  &  6.9488  &  3.6  &  -17.7  &  3.6  &  0.1990  &  0.0061  &  54.1  \\
57718.55149  &  -18.18923  &  1.73  &  6.9507  &  3.5  &  -12.9  &  3.5  &  0.2019  &  0.0060  &  56.2  \\
57719.55290  &  -18.19238  &  1.71  &  6.9511  &  3.4  &  -22.8  &  3.4  &  0.1884  &  0.0056  &  56.5  \\
57719.57368  &  -18.19477  &  1.71  &  6.9425  &  3.4  &  -18.4  &  3.4  &  0.1994  &  0.0057  &  56.7  \\
57720.53108  &  -18.18721  &  1.45  &  6.9469  &  2.9  &  -14.2  &  2.9  &  0.1922  &  0.0046  &  70.4  \\
57720.55102  &  -18.18671  &  1.52  &  6.9397  &  3.0  &  -17.1  &  3.0  &  0.1968  &  0.0051  &  65.7  \\
57721.53077  &  -18.19020  &  2.01  &  6.9556  &  4.0  &  -24.1  &  4.0  &  0.1938  &  0.0070  &  47.5  \\
57721.55300  &  -18.18965  &  2.09  &  6.9408  &  4.2  &  -14.9  &  4.2  &  0.1838  &  0.0076  &  45.7  \\
57935.79544  &  -18.17258  &  2.34  &  6.9898  &  4.7  &  -7.0  &  4.7  &  0.2700  &  0.0091  &  42.0  \\
57935.81684  &  -18.17136  &  2.21  &  6.9761  &  4.4  &  -5.1  &  4.4  &  0.2564  &  0.0084  &  44.2  \\
57936.84590  &  -18.17932  &  2.48  &  6.9803  &  5.0  &  -15.5  &  5.0  &  0.2820  &  0.0100  &  39.7  \\
57936.86711  &  -18.18014  &  2.53  &  6.9722  &  5.1  &  1.8  &  5.1  &  0.2668  &  0.0104  &  39.1  \\
57937.77515  &  -18.17689  &  2.58  &  6.9604  &  5.2  &  -6.9  &  5.2  &  0.2415  &  0.0105  &  38.3  \\
57937.82206  &  -18.17986  &  2.30  &  6.9765  &  4.6  &  -15.1  &  4.6  &  0.2444  &  0.0088  &  42.4  \\
57942.77873  &  -18.18570  &  1.58  &  6.9310  &  3.2  &  -19.1  &  3.2  &  0.2124  &  0.0046  &  61.4  \\
57942.88776  &  -18.19012  &  1.65  &  6.9373  &  3.3  &  -15.1  &  3.3  &  0.2270  &  0.0066  &  61.1  \\
57943.75116  &  -18.18923  &  1.62  &  6.9337  &  3.2  &  -13.2  &  3.2  &  0.1961  &  0.0048  &  60.4  \\
57943.86160  &  -18.19289  &  2.46  &  6.9371  &  4.9  &  -21.2  &  4.9  &  0.1863  &  0.0097  &  39.7  \\
57944.77832  &  -18.19228  &  2.24  &  6.9244  &  4.5  &  -15.5  &  4.5  &  0.1864  &  0.0082  &  42.9  \\
57944.86188  &  -18.19112  &  1.99  &  6.9344  &  4.0  &  -25.0  &  4.0  &  0.1829  &  0.0075  &  48.7  \\
57945.75844  &  -18.18714  &  1.79  &  6.9247  &  3.6  &  -23.8  &  3.6  &  0.1759  &  0.0057  &  53.7  \\
57946.77897  &  -18.19034  &  2.30  &  6.9292  &  4.6  &  -25.4  &  4.6  &  0.1791  &  0.0080  &  42.0  \\
57948.80722  &  -18.18963  &  2.00  &  6.9241  &  4.0  &  -25.8  &  4.0  &  0.1690  &  0.0069  &  48.4  \\
57948.86573  &  -18.19051  &  2.09  &  6.9274  &  4.2  &  -21.6  &  4.2  &  0.1760  &  0.0080  &  46.5  \\
57949.84184  &  -18.18010  &  5.61  &  6.9040  &  11.2  &  -11.5  &  11.2  &  0.1322  &  0.0280  &  20.3  \\
57951.76491  &  -18.19084  &  4.40  &  6.9380  &  8.8  &  -23.9  &  8.8  &  0.2072  &  0.0226  &  24.4  \\
57951.85922  &  -18.17271  &  3.38  &  6.9374  &  6.8  &  -17.4  &  6.8  &  0.1802  &  0.0169  &  30.3  \\
57952.73957  &  -18.17978  &  3.26  &  6.9507  &  6.5  &  -13.5  &  6.5  &  0.1720  &  0.0148  &  31.2  \\
57952.86046  &  -18.17574  &  2.49  &  6.9466  &  5.0  &  -27.4  &  5.0  &  0.1881  &  0.0111  &  39.7  \\
57953.85555  &  -18.18618  &  2.46  &  6.9581  &  4.9  &  -19.0  &  4.9  &  0.1945  &  0.0105  &  40.1  \\
57954.82099  &  -18.17750  &  3.16  &  6.9479  &  6.3  &  -23.7  &  6.3  &  0.1541  &  0.0150  &  32.5  \\
57955.75309  &  -18.17132  &  2.30  &  6.9634  &  4.6  &  -16.6  &  4.6  &  0.1838  &  0.0080  &  41.8  \\
57955.91602  &  -18.17448  &  1.72  &  6.9752  &  3.4  &  -26.9  &  3.4  &  0.1939  &  0.0077  &  59.6  \\
57956.72888  &  -18.17041  &  2.32  &  6.9676  &  4.6  &  -25.0  &  4.6  &  0.2022  &  0.0082  &  41.8  \\
57956.91804  &  -18.17142  &  2.42  &  6.9667  &  4.8  &  -33.5  &  4.8  &  0.2255  &  0.0106  &  41.0  \\
57957.89416  &  -18.17995  &  4.32  &  6.9690  &  8.6  &  -10.7  &  8.6  &  0.1634  &  0.0218  &  25.0  \\
57959.78548  &  -18.17701  &  2.51  &  6.9612  &  5.0  &  -15.7  &  5.0  &  0.2194  &  0.0103  &  39.4  \\
57959.90585  &  -18.17966  &  1.91  &  6.9555  &  3.8  &  -16.3  &  3.8  &  0.2075  &  0.0084  &  52.1  \\
57960.74725  &  -18.18476  &  2.68  &  6.9651  &  5.4  &  -20.1  &  5.4  &  0.2090  &  0.0104  &  36.9  \\
57960.84628  &  -18.18490  &  2.65  &  6.9605  &  5.3  &  -12.9  &  5.3  &  0.1981  &  0.0108  &  37.4  \\
57961.76442  &  -18.17957  &  2.22  &  6.9582  &  4.4  &  -7.1  &  4.4  &  0.2149  &  0.0087  &  44.4  \\
57961.83425  &  -18.18436  &  2.41  &  6.9526  &  4.8  &  -16.7  &  4.8  &  0.2094  &  0.0171  &  43.3  \\
57962.78393  &  -18.17748  &  1.96  &  6.9541  &  3.9  &  -2.2  &  3.9  &  0.2119  &  0.0081  &  51.1  \\
57962.87271  &  -18.19164  &  5.55  &  6.9251  &  11.1  &  -18.0  &  11.1  &  0.2266  &  0.0438  &  21.7  \\
57964.74231  &  -18.17793  &  1.84  &  6.9650  &  3.7  &  -11.6  &  3.7  &  0.2149  &  0.0061  &  52.7  \\
57964.83225  &  -18.18182  &  1.94  &  6.9492  &  3.9  &  -20.0  &  3.9  &  0.2078  &  0.0089  &  51.7  \\
57965.73570  &  -18.19438  &  5.34  &  6.9277  &  10.7  &  -25.5  &  10.7  &  0.1945  &  0.0308  &  21.4  \\
57965.83742  &  -18.18131  &  5.94  &  6.9409  &  11.9  &  -8.8  &  11.9  &  0.2740  &  0.0440  &  20.4  \\
57993.66968  &  -18.18228  &  2.03  &  6.9482  &  4.1  &  -22.7  &  4.1  &  0.2104  &  0.0075  &  47.0  \\
57993.78491  &  -18.17993  &  2.08  &  6.9523  &  4.2  &  -18.3  &  4.2  &  0.1954  &  0.0086  &  46.3  \\
57993.84799  &  -18.18314  &  1.83  &  6.9520  &  3.7  &  -27.1  &  3.7  &  0.1934  &  0.0088  &  53.9  \\
57994.63060  &  -18.17385  &  1.98  &  6.9446  &  4.0  &  -22.7  &  4.0  &  0.1980  &  0.0070  &  47.8  \\
57994.74383  &  -18.17639  &  2.22  &  6.9466  &  4.4  &  -16.1  &  4.4  &  0.1864  &  0.0091  &  43.3  \\
57994.82005  &  -18.17586  &  2.19  &  6.9553  &  4.4  &  -23.9  &  4.4  &  0.2004  &  0.0102  &  44.6  \\
57995.63098  &  -18.17764  &  3.46  &  6.9558  &  6.9  &  -23.8  &  6.9  &  0.1701  &  0.0174  &  29.9  \\
57998.62754  &  -18.19046  &  1.56  &  6.9459  &  3.1  &  -15.7  &  3.1  &  0.1987  &  0.0046  &  61.3  \\
57998.71879  &  -18.19124  &  2.02  &  6.9506  &  4.0  &  -14.9  &  4.0  &  0.1797  &  0.0087  &  48.3  \\
57998.81073  &  -18.18741  &  2.21  &  6.9488  &  4.4  &  -17.8  &  4.4  &  0.1645  &  0.0116  &  45.0  \\
58008.66469  &  -18.18777  &  3.48  &  6.9314  &  7.0  &  -7.3  &  7.0  &  0.1958  &  0.0153  &  29.0  \\
58010.65231  &  -18.19970  &  2.19  &  6.9277  &  4.4  &  -28.0  &  4.4  &  0.1689  &  0.0085  &  43.3  \\
58010.77942  &  -18.19305  &  2.25  &  6.9234  &  4.5  &  -26.8  &  4.5  &  0.1921  &  0.0107  &  43.3  \\
58010.83344  &  -18.19911  &  2.07  &  6.9391  &  4.1  &  -28.7  &  4.1  &  0.1985  &  0.0115  &  47.9  \\
58011.68577  &  -18.18190  &  2.40  &  6.9283  &  4.8  &  -9.4  &  4.8  &  0.1728  &  0.0105  &  40.2  \\
58011.77789  &  -18.18417  &  2.23  &  6.9321  &  4.5  &  -15.3  &  4.5  &  0.1739  &  0.0104  &  43.6  \\
58011.83205  &  -18.18652  &  2.60  &  6.9261  &  5.2  &  -37.8  &  5.2  &  0.1888  &  0.0150  &  38.6  \\
58012.67262  &  -18.18860  &  2.91  &  6.9347  &  5.8  &  -29.1  &  5.8  &  0.1680  &  0.0137  &  34.0  \\
58012.75336  &  -18.19386  &  2.31  &  6.9312  &  4.6  &  -35.4  &  4.6  &  0.1780  &  0.0106  &  42.0  \\
58012.82174  &  -18.18669  &  2.43  &  6.9230  &  4.9  &  -17.5  &  4.9  &  0.1579  &  0.0132  &  40.9  \\
58013.67492  &  -18.18197  &  3.38  &  6.9334  &  6.8  &  -28.1  &  6.8  &  0.1733  &  0.0158  &  30.1  \\
58013.74342  &  -18.18162  &  2.52  &  6.9295  &  5.0  &  -21.7  &  5.0  &  0.1862  &  0.0109  &  38.5  \\
58013.81433  &  -18.18331  &  2.17  &  6.9467  &  4.3  &  -31.5  &  4.3  &  0.1964  &  0.0106  &  45.2  \\
58014.68555  &  -18.18273  &  2.62  &  6.9207  &  5.2  &  -26.3  &  5.2  &  0.2481  &  0.0102  &  37.1  \\
58014.77534  &  -18.18936  &  2.77  &  6.9302  &  5.5  &  -28.6  &  5.5  &  0.1661  &  0.0130  &  36.0  \\
58018.63224  &  -18.18640  &  2.33  &  6.9479  &  4.7  &  -28.0  &  4.7  &  0.1890  &  0.0095  &  40.9  \\
58018.72534  &  -18.18438  &  2.52  &  6.9262  &  5.0  &  -23.1  &  5.0  &  0.1978  &  0.0108  &  38.2  \\
58019.62715  &  -18.19033  &  1.99  &  6.9432  &  4.0  &  -20.6  &  4.0  &  0.1888  &  0.0077  &  47.8  \\
58019.72843  &  -18.19547  &  1.92  &  6.9260  &  3.8  &  -25.9  &  3.8  &  0.1963  &  0.0080  &  49.9  \\
58020.66431  &  -18.18766  &  3.01  &  6.9468  &  6.0  &  -5.8  &  6.0  &  0.1870  &  0.0121  &  32.5  \\
58020.75915  &  -18.18656  &  2.21  &  6.9431  &  4.4  &  -19.1  &  4.4  &  0.1977  &  0.0099  &  43.6  \\
58021.55154  &  -18.17666  &  2.55  &  6.9344  &  5.1  &  -19.9  &  5.1  &  0.2109  &  0.0103  &  37.6  \\
58021.66021  &  -18.18003  &  3.07  &  6.9414  &  6.1  &  -35.5  &  6.1  &  0.2225  &  0.0132  &  32.3  \\
58021.75738  &  -18.18448  &  2.24  &  6.9325  &  4.5  &  -15.9  &  4.5  &  0.1845  &  0.0100  &  43.1  \\
58022.53363  &  -18.18830  &  2.59  &  6.9472  &  5.2  &  -24.9  &  5.2  &  0.1947  &  0.0103  &  37.1  \\
58022.59935  &  -18.18210  &  3.26  &  6.9427  &  6.5  &  -17.1  &  6.5  &  0.1937  &  0.0128  &  30.2  \\
58022.74556  &  -18.19200  &  2.80  &  6.9414  &  5.6  &  -22.3  &  5.6  &  0.2040  &  0.0124  &  35.0  \\
58023.53440  &  -18.17930  &  1.81  &  6.9489  &  3.6  &  -32.0  &  3.6  &  0.2030  &  0.0058  &  51.1  \\
58023.63020  &  -18.17708  &  2.62  &  6.9523  &  5.2  &  -19.1  &  5.2  &  0.1963  &  0.0106  &  36.7  \\
58023.75632  &  -18.18220  &  2.00  &  6.9325  &  4.0  &  -24.1  &  4.0  &  0.1824  &  0.0092  &  48.3  \\
58025.54957  &  -18.17300  &  1.67  &  6.9443  &  3.3  &  -17.9  &  3.3  &  0.2031  &  0.0055  &  56.1  \\
58025.64932  &  -18.17642  &  1.82  &  6.9560  &  3.6  &  -19.6  &  3.6  &  0.1990  &  0.0074  &  52.8  \\
58025.70429  &  -18.17085  &  1.93  &  6.9569  &  3.9  &  -4.2  &  3.9  &  0.2057  &  0.0089  &  50.3  \\
58026.55716  &  -18.16496  &  4.02  &  6.9580  &  8.0  &  -8.4  &  8.0  &  0.2396  &  0.0170  &  25.7  \\
58026.65617  &  -18.17836  &  2.73  &  6.9569  &  5.5  &  -22.0  &  5.5  &  0.2263  &  0.0112  &  35.7  \\
58026.74871  &  -18.18277  &  2.88  &  6.9711  &  5.8  &  -22.1  &  5.8  &  0.1771  &  0.0154  &  35.2  \\
58027.56593  &  -18.15614  &  5.14  &  6.9831  &  10.3  &  -11.8  &  10.3  &  0.2023  &  0.0274  &  21.8  \\
58027.67900  &  -18.16790  &  4.30  &  7.0036  &  8.6  &  -8.0  &  8.6  &  0.1720  &  0.0246  &  25.4  \\
58027.74799  &  -18.15559  &  5.45  &  6.9669  &  10.9  &  28.2  &  10.9  &  0.1827  &  0.0322  &  21.1  \\
58041.54788  &  -18.19556  &  2.00  &  6.9314  &  4.0  &  -31.2  &  4.0  &  0.1730  &  0.0067  &  47.1  \\
58043.56762  &  -18.18878  &  1.35  &  6.9287  &  2.7  &  -24.3  &  2.7  &  0.2094  &  0.0039  &  73.6  \\
58043.72519  &  -18.18693  &  1.89  &  6.9299  &  3.8  &  -25.9  &  3.8  &  0.2145  &  0.0086  &  52.1  \\
58052.55248  &  -18.18881  &  1.46  &  6.9380  &  2.9  &  -22.7  &  2.9  &  0.2180  &  0.0046  &  66.9  \\
58052.61595  &  -18.19051  &  1.95  &  6.9391  &  3.9  &  -27.6  &  3.9  &  0.2350  &  0.0085  &  50.3  \\
58053.56897  &  -18.18273  &  1.70  &  6.9377  &  3.4  &  -27.6  &  3.4  &  0.2239  &  0.0060  &  56.8  \\
58053.64640  &  -18.18326  &  1.80  &  6.9436  &  3.6  &  -23.4  &  3.6  &  0.2202  &  0.0074  &  54.6  \\
58054.52491  &  -18.18453  &  2.44  &  6.9497  &  4.9  &  -36.6  &  4.9  &  0.2112  &  0.0083  &  39.1  \\
58054.67331  &  -18.18235  &  2.09  &  6.9612  &  4.2  &  -21.0  &  4.2  &  0.2261  &  0.0086  &  46.9  \\
58056.53663  &  -18.18039  &  1.35  &  6.9497  &  2.7  &  -23.1  &  2.7  &  0.2132  &  0.0039  &  74.8  \\
58056.62290  &  -18.18640  &  1.60  &  6.9600  &  3.2  &  -33.6  &  3.2  &  0.2198  &  0.0064  &  62.6  \\
58057.53537  &  -18.18602  &  1.32  &  6.9511  &  2.6  &  -28.4  &  2.6  &  0.2202  &  0.0038  &  76.5  \\
58057.59482  &  -18.18711  &  1.41  &  6.9528  &  2.8  &  -29.3  &  2.8  &  0.2291  &  0.0049  &  71.7  \\
58068.58875  &  -18.18136  &  1.63  &  6.9330  &  3.3  &  -14.5  &  3.3  &  0.1845  &  0.0055  &  59.0  \\
58069.65402  &  -18.19329  &  1.90  &  6.9399  &  3.8  &  -17.5  &  3.8  &  0.1783  &  0.0062  &  49.6  \\
58070.66278  &  -18.19205  &  1.93  &  6.9408  &  3.9  &  -20.7  &  3.9  &  0.1868  &  0.0086  &  50.8  \\
58071.54887  &  -18.18911  &  1.69  &  6.9187  &  3.4  &  -23.1  &  3.4  &  0.1924  &  0.0061  &  56.8  \\
58074.55346  &  -18.18790  &  1.55  &  6.9236  &  3.1  &  -18.4  &  3.1  &  0.1788  &  0.0060  &  62.4  \\
58075.53852  &  -18.18124  &  1.76  &  6.9243  &  3.5  &  -18.9  &  3.5  &  0.1630  &  0.0069  &  54.5  \\
58077.54947  &  -18.18492  &  1.60  &  6.9297  &  3.2  &  -23.9  &  3.2  &  0.1872  &  0.0063  &  61.9  \\
58079.61835  &  -18.19371  &  1.94  &  6.9281  &  3.9  &  -21.6  &  3.9  &  0.1803  &  0.0084  &  49.5  \\
\end{longtable}

\newpage

\begin{landscape}
\begin{longtable}{lccc}
\caption{List of parameters used in the analysis. The respective priors are provided together with the posteriors for both the Dartmouth and PARSEC stellar evolution tracks. The posterior values represent the median and 68.3\% credible interval. Fixed and derived values that might be useful for follow-up work are also reported. \label{MCMCprior}}\\
\hline
Parameter & Prior & \multicolumn{2}{c}{Posterior}\\
 &  & Dartmouth & PARSEC\\
 &  & (adopted) & \\
\hline
\endfirsthead
\multicolumn{4}{l}{{\bfseries \tablename\ \thetable{} -- continued from previous page}} \\
\hline
Parameter & Prior & \multicolumn{2}{c}{Posterior}\\
 &  & Dartmouth & PARSEC\\
\hline
\endhead
\multicolumn{4}{l}{{Continued on next page}} \\ 
\hline
\endfoot
\hline
\multicolumn{4}{l}{Notes:}\\
\multicolumn{4}{l}{$\bullet$ $\mathcal{N}(\mu,\sigma^{2})$: normal distribution with mean $\mu$ and width $\sigma^{2}$}\\
\multicolumn{4}{l}{$\bullet$ $\mathcal{U}(a,b)$: uniform distribution between $a$ and $b$}\\
\multicolumn{4}{l}{$\bullet$ $\mathcal{N}_{\mathcal{U}}(\mu,\sigma^{2},a,b)$: normal distribution with mean $\mu$ and width $\sigma^{2}$ multiplied with a uniform distribution between $a$ and $b$}\\
\multicolumn{4}{l}{$\bullet$ $\mathcal{S}(a,b)$: sine distribution between $a$ and $b$}\\
\multicolumn{4}{l}{$\bullet$ $\beta(a,b)$: Beta distribution with parameters $a$ and $b$}\\
\endlastfoot
\multicolumn{4}{l}{\it Stellar Parameters}\\
Effective temperature \teff\ [K] & $\mathcal{N}(5457,29)$ & $5477\pm27$ & $5480\pm24$\\
Surface gravity \logg\ [cgs] & $\mathcal{N}(4.42,0.10)$ & $4.419\pm0.053$ & $4.429\pm0.045$\\
Iron abundance \met\ [dex] &  $\mathcal{N}(0.08,0.02)$ & $0.078\pm0.020$ & $0.079\pm0.020$\\
Distance to Earth $D$ [pc] &  $\mathcal{N}(143.5,10.9)$ & $145\pm8$ & $141\pm6$\\
Interstellar extinction $E(B-V)$ [mag] &  $\mathcal{U}(0,1)$ & $0.009^{_{+0.011}}_{^{-0.007}}$ & $0.009^{_{+0.011}}_{^{-0.007}}$\\
Systemic radial velocity $\gamma$ [\kms] & $\mathcal{U}(-20,-15)$ & $-18.186\pm0.002$ & $-18.186\pm0.002$\\
Linear limb-darkening coefficient $u_{a}$ & (derived) & $0.4631\pm0.0061$ & $0.4625\pm0.0057$ \\
Quadratic limb-darkening coefficient $u_{b}$ & (derived) & $0.2270\pm0.0041$ & $0.2273\pm0.0037$ \\
Stellar density $\rho_{\star}/\rho_{\astrosun}$ & (derived) & $0.98\pm0.19$ & $1.03\pm0.16$ \\
Stellar mass M$_{\star}$\ [\Msun] & (derived) & $0.915\pm0.017$ & $0.884\pm0.018$\\
Stellar radius R$_{\star}$\ [\Rsun] & (derived) & $0.977\pm0.053$ & $0.950\pm0.040$\\
Stellar age $\tau$\ [Gyr] & (derived) & $9.7\pm3.0$ & $10.8\pm2.8$\\
&&& \\
\hline
\multicolumn{4}{l}{\it Planet b Parameters}\\
Orbital Period $P$ [d] & $\mathcal{N}(2.369193,0.01)$ & $2.369172\pm8.9\times10^{-5}$ & $2.369173\pm9.0\times10^{-5}$ \\
Transit epoch $T_{0}$ [BJD - 2456000] & $\mathcal{N}(981.6425,0.1)$ & $981.6431\pm1.6\times10^{-3}$ & $981.6431\pm1.6\times10^{-3}$\\
Radial velocity semi-amplitude $K$ [\ms] & $\mathcal{U}(0,10^{2})$ & $3.34\pm0.43$ & $3.33\pm0.43$\\
Orbital inclination $i$ [$^{\circ}$] & $\mathcal{S}(70,90)$ & $87.7\pm1.6$ & $88.1\pm1.4$\\
Planet-to-star radius ratio $k$ & $\mathcal{U}(0,1)$ & $0.01604\pm0.00041$ & $0.01599\pm0.00035$\\
Orbital eccentricity $e$ & $\mathcal{U}(0,1)$ & $0.084\pm0.079$ & $0.080\pm0.068$ \\
Argument of periastron $\omega$ [\degr] & $\mathcal{U}(0,360)$ & $99^{_{+220}}_{^{-77}}$ & $94^{_{+220}}_{^{-71}}$ \\
System scale $a/R_{\star}$ & (derived) & $7.43\pm0.45$ & $7.56\pm3.8$\\
Impact parameter $b$ & (derived) & $0.30\pm0.20$ & $0.25\pm0.19$\\
Transit duration T$_{14}$ [h] & (derived) & $2.266\pm0.050$ & $2.264\pm0.049$\\
Semi-major axis $a$ [AU] & (derived) & $0.03376\pm0.00021$ & $0.03337\pm0.00023$\\
Planet mass M$_{p}$ [\Mearth] &(derived) & $6.54 \pm 0.84$ & $6.38 \pm 0.83$\\
Planet radius R$_{p}$ [\Rearth] & (derived) & $1.71\pm0.11$ & $1.654\pm0.84$\\
Planet bulk density $\rho_{p}$ [\gcm3] & (derived) & $7.1\pm1.8$ & $7.7\pm1.7$\\
&&& \\
\hline
\multicolumn{4}{l}{\it Gaussian Process Hyperparameters}\\
$A$ [\ms] & $\mathcal{U}(0,100)$ & $6.0^{_{+1.3}}_{^{-0.8}}$ & $5.96\pm1.2$\\
$\lambda_{1}$ [d] & $\mathcal{U}(0,100)$ & $34\pm12$ & $34\pm12$\\
$\lambda_{2}$ & $\mathcal{U}(0,10)$ & $0.46\pm0.12$ & $0.46\pm0.12$\\
$P_{rot}$ [d] & $\mathcal{N}(32.2,0.6)$ & $32.2\pm0.5$ & $32.2\pm0.5$\\
&&& \\
\hline
\multicolumn{4}{l}{\it Instrument-related Parameters}\\
HARPS jitter [\ms] & $\mathcal{U}(0,10^{2})$ & $1.9\pm0.4$ & $1.9\pm0.4$\\
\textit{K2} contamination [\%] & $\mathcal{N_{U}}(0.952,0.024,0,100)$ & $0.952\pm0.024$ & $0.952\pm0.024$\\
\textit{K2} jitter [ppm] & $\mathcal{U}(0,10^{5})$ & $59\pm1$ & $59\pm1$\\
\textit{K2} out-of-transit flux & $\mathcal{U}(0.99,1.01)$ & $1.000006\pm2\times10^{-6}$ & $1.000006\pm2\times10^{-6}$\\
SED jitter [mag] & $\mathcal{U}(0,0.1)$ & $0.054\pm0.021$ & $0.054\pm0.022$\\
\end{longtable}
\end{landscape}
\end{appendix}

\end{document}